\documentclass[12pt]{article}

\newlength{\abstractwidth}
\usepackage{cancel}
\usepackage{hyperref}
\usepackage{color}
\usepackage{graphicx}
\usepackage{tikz}
\usepackage{mathtools}
\usepackage{float}

\DeclarePairedDelimiter\floor{\lfloor}{\rfloor}

\usepackage{cancel}
\usepackage{hyperref}

\usepackage{cite}
\usepackage{amsmath}
\usepackage{amssymb}
\usepackage{latexsym}
 \usepackage{setspace}
 
 \usepackage{caption}
\usepackage{subcaption}

\numberwithin{equation}{section}

\newcommand{\abs}[1]{\left\lvert #1 \right\rvert}
\renewcommand{\thefootnote}{\fnsymbol{footnote}}
\renewcommand{\thanks}[1]{\footnote{#1}}
\newcommand{\starttext}{
\setcounter{footnote}{0}
\renewcommand{\thefootnote}{\arabic{footnote}}}
\newcommand{\bea}{\begin{eqnarray}}
\newcommand{\eea}{\end{eqnarray}}
\newcommand{\be}{\begin{eqnarray}}
\newcommand{\ee}{\end{eqnarray}}

\def\ie{\begin{equation}\begin{aligned}}
\def\fe{\end{aligned}\end{equation}}
\def\half{{\scriptstyle \frac 12}}

\def\sevenh{{\scriptstyle \frac 72}}
\def\threeh{{\scriptstyle \frac 32}}
\def\fiveh{{\scriptstyle \frac 52}}

\def\nineh{{\scriptstyle \frac 92}}

\def\ie{\begin{equation}\begin{aligned}}
\def\fe{\end{aligned}\end{equation}}

\def\cB{{\cal B}}
\def\cC{{\cal C}}

\def\cE{{\cal E}}

\def\cH{{\cal H}}

\def\cN{{\cal N}}
\def\cO{{\cal O}}

\def\cQ{{\cal Q}}

\def\cS{{\cal S}}
\def\cT{{\cal T}}

\def\ZZ{{\mathbb Z}}

\def\NN{{\mathbb N}}

\def\nn{\nonumber}
\def\Re{{\rm Re \,}}
\def\Im{{\rm Im \,}}

\def\p{\partial}

\def\stau{\tau}

\def\ttau{\rho}

\def\Z{{\mathbb Z}}

\def\sint{{ s}}
\def\tint{{ t}}

\def\bt{b.t.}

\newcommand{\btau}{}

\newcommand{\oone}{\omega_1}
\newcommand{\otwo}{\omega_2}

\newcommand{\es}[2] {\begin{equation} \label{#1} \begin{split} #2 \end{split} \end{equation}}

\begin{document}
\begin{flushright}
	QMUL-PH-23-20
\end{flushright}

\starttext

\setcounter{footnote}{0}

\vskip 0.3in

\centerline{\large \bf Relations between integrated correlators in} 
\vskip 0.1in
\centerline{\large \bf $\cN=4$  Supersymmetric Yang--Mills Theory}

\vskip 0.2in

\vskip 0.2in
\noindent Luis F. Alday$^{1}$,  Shai M. Chester$^{2,3,4}$, Daniele Dorigoni$^{5}$, Michael B. Green$^6$, Congkao Wen$^{7}$
\vskip 0.15in

\vskip 0.1in

\noindent{\small $^1$ Mathematical Institute,  University of Oxford,
  Woodstock Road, Oxford OX2 6GG, UK}

\noindent\vskip 0.03in
\noindent{\small $^2$ {Jefferson Physical Laboratory, Harvard University,}
 Cambridge, MA 02138, USA}
 
\vskip 0.03in

\noindent{\small $^3$ {CMSA, Harvard University, Cambridge, MA 02138, USA}}

\vskip 0.03in
\noindent{\small $^4$ {Blackett Laboratory, Imperial College, Prince Consort Road, London, SW7 2AZ, UK}}

\vskip 0.03in

\noindent{\small $^5$ {Centre for Particle Theory, Durham University, Stockton Road, Durham DH1 3LE, UK}}

\vskip 0.03in

\noindent{\small $^6$ {DAMTP,   Wilberforce Road, Cambridge CB3 0WA, UK}}

\vskip 0.03in

\noindent{\small  $^7$ {Centre for Theoretical Physics, Queen Mary University of London,  London, E1 4NS, UK}}

\vskip 0.5in

\begin{abstract}
\vskip 0.1in

Integrated correlation functions in $\cN=4$ supersymmetric Yang--Mills theory with gauge group $SU(N)$ can be expressed in terms of the localised  $S^4$  partition function, $Z_N$, deformed by a mass $m$. Two such cases are $\cC_N=(\Im \tau)^2 \partial_\tau\partial_{\bar\tau} \partial_m^2\log Z_N\vert_{m=0}$  and $\cH_N=\partial_m^4\log Z_N\vert_{m=0}$, which are modular invariant functions of the complex coupling $\tau$. 
While $\cC_N$ was recently written in terms of a two-dimensional lattice sum for any $N$ and $\tau$, $\cH_N$  has only  been evaluated up to order $1/N^3$ in a large-$N$ expansion in terms of modular invariant functions with no known lattice sum realisation. Here we develop methods for evaluating  $\cH_N$ to any desired order in $1/N$ and finite $\tau$. We use this new data to constrain higher loop corrections to the stress tensor correlator, and give evidence for several intriguing relations between $\cH_N$ and $\cC_N$ to all orders in $1/N$. We also give evidence that the coefficients of the $1/N$ expansion of  $\cH_N$ can be written as lattice sums to all orders. Lastly, these large $N$ and finite $\tau$ results are used to accurately estimate the integrated correlators at finite $N$ and finite $\tau$. 

\end{abstract}      
                                   
\date{}

\newpage

\tableofcontents
\newpage

\section{Introduction}
\label{sec:introduction}

Four-dimensional maximally  supersymmetric Yang-Mills ($\cN=4$  SYM) theory with gauge group $SU(N)$ is of great interest for a variety of reasons.  It is a non-trivial four-dimensional superconformal gauge theory parameterised by the complex coupling $\tau\equiv \frac{\theta}{2\pi}+ i\frac{4\pi}{g_{_{YM}}^2}$ with Montonen--Olive duality group $SL(2,\ZZ)$ \cite{Montonen};\footnote{In this paper we will focus on the theory with gauge group $SU(N)$. S-duality involving more general gauge groups is discussed by Goddard, Nuyts and Olive in \cite{Goddard:1976qe}.} it is a toy model of QCD and an integrable system in the planar limit. Importantly for our present considerations it provides a  non-perturbative formulation of type IIB string theory in an $AdS_5\times S^5$ background \cite{Maldacena:1997re}\footnote{Similar holographic dualities exist for other gauge groups \cite{Witten:1998xy}.}.  In this holographic interpretation, type IIB string theory with coupling $\tau_s=\chi+i/g_s$ and string length scale $\ell_s$  in a background $AdS_5\times S^5$  of length scale $L$ is identified with $\cN=4$ SYM with gauge group $SU(N)$, where $\tau_s=\tau$ and $(L/\ell_s)^4= g_{_{YM}}^2 N$.
Therefore, the gravity description at small $\ell_s$ is dual to SYM at large $N$ and finite $\tau$.\footnote{At large $N$ and finite $\tau$, the bulk is weakly curved because $\ell_s$ is small, but the string theory is still strongly coupled because $\tau_s$ is finite. One can further take $g_s$ to be weakly coupled, which is then dual to the familiar large-$N$ large-$\lambda\equiv g_{_{YM}}^2N$ limit of SYM.} Super-graviton scattering amplitudes in string theory on $AdS_5\times S^5$ are then dual to correlation functions of the stress tensor multiplet. While this duality in principle gives a non-perturbative definition of string theory in terms of a well defined CFT, in practice it is difficult to use the duality to study string theory in the gravity regime since the CFT is strongly coupled at large $N$. We thus need to study $\cN=4$ SYM non-perturbatively to make progress in  studying quantum gravity via AdS/CFT.

\subsection{Brief review of integrated correlators}
\label{sec:review}

Certain properties of $\cN=4$ SYM can be obtained analytically for all values of $N$ and $\tau$ using supersymmetric localisation (see \cite{Pestun:2016zxk} for a review). This method was originally used to compute non-local quantities such as supersymmetric partition functions and Wilson loops \cite{Pestun:2007rz}.  
More recently,  it was understood \cite{Binder:2019jwn,Chester:2020dja} that certain integrals of the correlator of four superconformal primary operators in the $\cN=4$ stress tensor multiplet are obtained from the following derivatives of the $S^4$ partition function $Z_N(m,\tau\btau)$ deformed by the $\mathcal{N}=2$ preserving mass $m$:
\es{eq:GHdef}{
\mathcal{C}_N(\tau\btau)\equiv \frac{1}{4}\Delta_\tau \partial_m^2\log Z_N(m,\tau\btau)\big\vert_{m=0}\,,\qquad \mathcal{H}_N(\tau\btau)&\equiv \partial_m^4\log Z_N(m,\tau\btau)\big\vert_{m=0}\,,
}
where $\Delta_\tau\equiv4\tau_2^2\partial_\tau\partial_{\bar\tau}$ is the hyperbolic laplacian and $\tau=\tau_1 + i \tau_2$ with $\tau_1={\theta}/{(2\pi)}$, and $\tau_2={4\pi}/{g_{_{YM}}^2}$. 
The quantities $\cC_N$ and $\cH_N$ are identified with integrals over  the insertion points of the operators in the four-point correlator.
The choice of integration measure distinguishes the choice of integrated correlators. 

The relations \eqref{eq:GHdef} are useful because $Z_N(m,\tau\btau)$ can be computed using supersymmetric localisation in terms of an $(N-1)$-dimensional matrix model integral \cite{Pestun:2007rz}. The integrand is a product of a classical contribution that depends on $g_{_{YM}}$, a one-loop contribution that is $\tau$-independent, and a sum over an infinite number of instanton contributions that depend on $\tau$ and $\bar\tau$ \cite{Moore:1997dj,Losev:1997tp,Nekrasov:2002qd,Nekrasov:2003rj}. In order to use these integrated correlators to study the gravity regime of string theory via AdS/CFT, we need to compute these $(N-1)$-dimensional integrals explicitly for large $N$.

The perturbative sector of $\mathcal{C}_N(\tau\btau)$ was first computed for both finite $N$ and in a large-$N$ expansion in \cite{Chester:2019pvm}. The instanton sectors were then also computed at large $N$ in \cite{Chester:2019jas}, and the large-$N$ expansion of  $\mathcal{C}_N(\tau\btau)$ was found to take the following form  
\es{eq:GN}{
\mathcal{C}_N(\tau\btau)&=\frac{ N^2}{4} -\frac{3\sqrt{N}}{2^4 \ } E( {\scriptstyle {3 \over 2}};\tau\btau)+\frac{45}{2^8 \sqrt{N}}E( {\scriptstyle {5 \over 2}};\tau\btau)+\frac{1}{{N}^{\frac32}}\left[-\frac{39}{2^{13} }E( {\scriptstyle {3 \over 2}};\tau\btau)+\frac{4725}{2^{15} }E( {\scriptstyle {7 \over 2}};\tau\btau)\right]\\
&\quad+\frac{1}{{N}^{\frac52}}\left[-\frac{1125}{2^{16} }E( {\scriptstyle {5 \over 2}};\tau\btau)+\frac{99225}{2^{18} }E( {\scriptstyle {9 \over 2}};\tau\btau)\right]+O(N^{-\frac{7}{2}})\,,
}
where $E(s;\tau\btau)$  is the  non-holomorphic Eisenstein series, that may be defined by the two-dimensional lattice sum\footnote{The normalisation of $E(s;\tau\btau)$ differs from that of \cite{Chester:2019jas} in a manner that eliminates factors of $\pi$ in the coefficients of the series in \eqref{eq:GN} and subsequent expressions.}
\be
E(s;\tau\btau) = \frac{1}{\pi^s}\sum_{(m,n)\,\neq\,(0,0)}\frac{\tau_2^s}{|m+n\tau|^{2s}}   \,,
\label{eq:EisensteinDef}
\ee
which is the unique $SL(2,\mathbb{Z})$ invariant solution of moderate growth to the homogeneous Laplace eigenvalue equation
	\es{LaplaceEq}{
		\left( \Delta_\tau - s(s-1) \right) E(s;\tau\btau) = 0 \, . } 
 
 It was shown in \cite{Dorigoni:2021bvj,Dorigoni:2021guq} that,	 for all $N$ and $\tau$, $\cC_N(\tau\btau)$ can be written as a two-dimensional lattice sum, 
  \begin{align}
\cC_{N} (\tau\btau)  =  \sum_{(m,n)\in\mathbb{Z}^2}  \int_0^\infty dt \, B_N(t)\exp\Big(- t \pi \frac{|m+n\tau|^2}{\tau_2} \Big) \,,
\label{eq:gsun}
\end{align} 
 where $B_N(t)= \cQ_N(t)/(t+1)^{2N+1}$ and $\cQ_N(t)$ is a 
 polynomial of degree $(2N+1)$.   Although the expression for $B_N(t)$ was arrived at in \cite{Dorigoni:2021guq}  by analysing the perturbative and 
instanton contributions for many values of $N$, a derivation of its form was given in \cite{Dorigoni:2022cua} based on the construction of a  generating series, $\cC(z;\tau\btau) \equiv \sum_{N=1}^\infty  \cC_N(\tau\btau) z^N$.  The expression  \eqref{eq:gsun} was shown to satisfy a Laplace difference equation
\es{eq:Grec}{
\Delta_\tau\cC_N(\tau\btau)=N^2\big[\cC_{N+1}(\tau\btau)-2\,\cC_N(\tau\btau)+\cC_{N-1}(\tau\btau)\big]-N\big[\cC_{N+1}(\tau\btau)-\cC_{N-1}(\tau\btau)\big]+2\,\cC_N(\tau\btau)\,.
}
Using the initial condition $\cC_1(\tau\btau)=0$, this recursion relation can be used to efficiently compute all $\cC_N(\tau\btau)$ in terms of $\cC_2(\tau\btau)$.

The $1/N$ expansion \eqref{eq:GN} is non-convergent, and the explicit modular invariant trans-series containing the large-$N$ completion was obtained by analysing the generating series \cite{Dorigoni:2022cua}. 
In the large-$\lambda$ limit, the large-$N$ completion behaves as $e^{-2\sqrt \lambda}$ as obtained first using resurgence \cite{Dorigoni:2021guq, Hatsuda:2022enx}. Many other results concerning the integrated correlator $\cC_N(\tau\btau)$ as well as its extensions for other gauge groups and more general 1/2-BPS operators can be found in the literature, see e.g. \cite{Alday:2021vfb,Dorigoni:2022zcr, Collier:2022emf,Brown:2023zbr,Brown:2023why,Brown:2023cpz,Wen:2022oky,Dorigoni:2021rdo,Green:2020eyj,Paul:2023rka,Paul:2022piq,Chester:2022sqb,Behan:2023fqq, Dorigoni:2023ezg}.

Much less is known about the second correlator $\cH_N(\tau)$, which is the focus of this paper. While the perturbative sector was computed at finite $N$ in \cite{Chester:2020dja}, the large-$N$ expansion of these perturbative terms is much more challenging and only the first few orders were computed in that study. In \cite{Chester:2020vyz}, it was shown that instanton terms could be computed more easily at large $N$, which were used to compute the first few terms at large $N$ and finite $\tau$.  The resulting series has the following form
\es{eq:HN}{
 &\cH_N(\tau\btau) =6 N^2+\cH^h_N(\tau\btau)+\cH^i_N(\tau\btau)+\dots\, ,
 }
 where ellipsis denotes terms that are exponentially suppressed in $N$. 
The series  $\cH^h_N$ contains terms with  half-integer powers of $1/N$, 
 \es{eq:HN2}{
& \cH^h_N(\tau\btau)=\sqrt{N} E( {\scriptstyle {3 \over 2}};\tau\btau)-\frac{9}{2\sqrt{N}}E( {\scriptstyle {5 \over 2}};\tau\btau)+\frac{1}{{N}^{\frac32}}\left[\frac{117}{2^8 }E( {\scriptstyle {3 \over 2}};\tau\btau)-\frac{3375}{2^{10} }E( {\scriptstyle {7 \over 2}};\tau\btau)\right]\\
&\quad+\frac{1}{{N}^{\frac52}}\left[\frac{675}{2^{10}}E( {\scriptstyle {5 \over 2}};\tau\btau)-\frac{33075}{2^{12} }E( {\scriptstyle {9 \over 2}};\tau\btau)\right]+O(N^{-\frac72})\,,\\
}
where the coefficients are rational multiples of non-holomorphic Eisenstein series, just as in the case of $\cC_N$ in \eqref{eq:GN}. The series $\cH^i_N$  contains integer powers of $1/N$.  The terms up to $O(1/N^3)$  were evaluated in  \cite{Chester:2020vyz} and take the following form
\begin{align}	\label{eq:HN3}
 &\cH^i_N(\tau\btau)=C_0+\frac{27}{2^3 N}\cE(4,{\scriptstyle {3 \over 2}},{\scriptstyle {3 \over 2}};\tau\btau)+\frac{1}{N^2}\left[C_1-\frac{14175}{704}\cE(7,{\scriptstyle {5 \over 2}},{\scriptstyle {3 \over 2}};\tau\btau)+\frac{1215}{88}\cE(5,{\scriptstyle {5 \over 2}},{\scriptstyle {3 \over 2}};\tau\btau)\right]\\
		+& \, \frac{1}{N^3}\Big[\alpha_3\cE(4,{\scriptstyle {3 \over 2}},{\scriptstyle {3 \over 2}};\tau\btau)+\sum_{s=6,8,9}[\alpha_s\cE(s,{\scriptstyle {3 \over 2}},{\scriptstyle {3 \over 2}};\tau\btau)+\beta_s\cE(s,{\scriptstyle {5 \over 2}},{\scriptstyle {5 \over 2}};\tau\btau)+\gamma_s\cE(s,{\scriptstyle {7 \over 2}},{\scriptstyle {3 \over 2}};\tau\btau)]\Big]\! +O(N^{-4})\,,\nn
\end{align}	
	where $\alpha_s,\beta_s,\gamma_s$ are known rational constants, while the values of the constants $C_0$ and $C_1$ require further knowledge of the perturbative sector. The function $\cE(s, s_1, s_2; \tau\btau)$ is a generalised Eisenstein series, which is an $SL(2,\mathbb{Z})$ invariant solution of the inhomogeneous Laplace eigenvalue equation
	\footnote{Note that in \cite{Chester:2020vyz}  a different convention was used in which
	$\hat {\cE}(s,s_1,s_2;\tau\btau)=- \pi^{s_1+s_2} \cE(s+1,s_1,s_2;\tau\btau)$, where  $\hat {\cE}(s,s_1,s_2;\tau\btau)$ denotes the generalised Eisenstein series of \cite{Chester:2020vyz}.}
		 \es{eq:laplacewsource}{
		\left( \Delta_\tau - s(s-1) \right) \cE(s, s_1, s_2; \tau\btau)=  E(s_1 ; \tau\btau)E(s_2; \tau\btau) \,.
} 
An example of such a function, $\cE(4, \threeh,\threeh; \tau\btau)$, entered as the coefficient of $d^6R^4$ in the low energy expansion of the type IIB effective action in \cite{Green:1999pu,Green:2005ba} where it was expressed as a four-dimensional lattice sum
 and its Fourier expansion was studied in \cite{Green:2014yxa}. More general examples of such lattice sums were obtained in \cite{Green:2008bf}
 \footnote{We will later see that such lattice sums require  more careful definitions than those given in \cite{Green:2005ba} and \cite{Green:2008bf}.}.  Starting at eigenvalue $s=6$ it was shown, in  \cite{Dorigoni:2021ngn} for cases when $s_1$ and $s_2$ are integers and in \cite{FKRinprogress} when $s_1$ and $s_2$ are half-integers, that  the Fourier-mode decomposition of $ \cE(s, s_1, s_2; \tau\btau)$ in general contains terms which are related to holomorphic cusp forms. This fact will play an important in our analysis. 
 
We will now summarise the main results of this paper, which will shed much more light on the structure of the large-$N$ expansion of $\cH_N$  and hint at its possible structure for all values of $N$.
 
\subsection{Summary of results}
\label{sec:summary}

Most of the explicit results in this paper apply to the large-$N$ expansion of $\cH_N$.  This involves developing techniques for determining the explicit form of the perturbative  terms in $\cH_N$ to any desired order in $1/N$.  When combined with the large-$N$ instanton expressions obtained following  \cite{Chester:2020vyz}, this determines  the full $\cH_N$ to any desired order in $1/N$.  

Several interesting relations between $\cH_N$ and $\cC_N$ will emerge from these explicit large-$N$ results. For example,  we find that the half-integer power terms, $\cH^h_N$ in \eqref{eq:HN} satisfy a Laplace difference equation 
\begin{align}
\Delta_\tau  {\cH_N^h(\tau\btau)} =&\label{eq:Hrec} \, N^2(N^2-1) \Big[\frac{\cH_{N+1}^h(\tau\btau)}{(N+1)^2}-2  {\cH_N^h(\tau\btau) \over N^2}  +\frac{\cH_{N-1}^h(\tau\btau)}{(N-1)^2} \Big]\\
&\nn+ \frac{16N}{N^2-1} \Big[\frac{3N^3}{4}-(N-1)^2 \cC_{N+1}(\tau\btau)+3 N\cC_N(\tau\btau)+(N+1)^2 \cC_{N-1}(\tau\btau)\Big]\,,
\end{align}
which allows us to efficiently compute $\cH^h_N$ to all orders in $1/N$. This recursion relation is similar to that of $\cC_N$ in \eqref{eq:Grec}, except now both integrated correlators appear on the right hand side, and furthermore we cannot use \eqref{eq:Hrec} to compute the full $\cH_N$ at finite $N$ because $\cH_N^i$ does not satisfy this recursion relation.

We also derived several new results concerning the integer power terms $\cH_N^i$. Firstly,  the $\tau$-independent terms $C_0$, $C_1$, $\dots$  that appear at even powers of $1/N$ in \eqref{eq:HN3} can now be evaluated. These coefficients can be used to constrain contact-term ambiguities of loop corrections to the holographic correlator. In particular, we will use the $C_0$ term to give a consistency check on the unique contact-term ambiguity of the one-loop graviton exchange term \cite{Alday:2017xua,Aprile:2017bgs}, which was originally fixed using $\cC_N$ in \cite{Chester:2019pvm}; while $C_1$ can be used to fix one of the four contact-term ambiguities of the 2-loop graviton exchange term \cite{Drummond:2022dxw,Huang:2021xws}. 

Although we suspect that the full integrated correlator, $\cH_N$, satisfies some differential equation analogous to the Laplace difference equation satisfied by $\cC_N$, this remains to be understood.  However, we find that the laplacian acting on $\cH_N^i$ is related to $\cC_N$ via a nonlinear relation  
\es{eq:Hrec2}{
\Delta_\tau \cH^i_N(\tau)\big\vert_E = \frac{96}{N^2-1}\, \left(\cC_N(\tau)-\frac{N^2}{4}\right)^2\,,
} 
where $\Delta_\tau \cH_N\vert_E$ means that the expression is restricted to the terms that are bilinear in non-holomorphic Eisenstein series  that appear when acting with the laplacian on the generalised Eisenstein series $\cE$, as appeared in the right-hand side of \eqref{eq:laplacewsource}. Although this relation  cannot recursively fix each order of $\cH_N^i$ as \eqref{eq:Hrec} did for $\cH_N^h$, we will see that it does strongly restrict the space of $\cE$ that arise in $\cH^i_N$ at each order in $1/N$ expansion.


As we commented earlier, an individual generalised Eisenstein series $\cE(s,s_1,s_2; \tau\btau)$ with eigenvalue $s\ge 6$ in general contains contributions which originate from holomorphic cusp forms of modular weight $2s$. However, we will show that the combinations of  $\cE(s,s_1,s_2; \tau\btau)$  that appear at each order of $1/N$ in \eqref{eq:HN3} and its extension to higher orders are very special ones in which the holomorphic cusp forms contributions cancel.  Closely related to this, the  combinations appearing in the $1/N$ expansion are precisely the combinations of generalised Eisenstein series that admit four-dimensional lattice sum representations. 
 
 More precisely we will see that at each order in $1/N$ up to order $O(N^{-7})$ that we have checked explicitly, the combination of generalised Eisenstein series that appears has the form of an integral of a four-dimensional lattice sum of the type that has appeared in \cite{Green:2008bf,DHoker:2018mys,Green:2014yxa} in the  study of the low energy expansion of type IIB string amplitudes.  Such specific sums of generalised Eisenstein series (modulo linear combinations of single Eisenstein series) have the form
\begin{align}
\cE^w_{i,j}(\tau\btau) =\!  \sum_{\underset {p_1+p_2+p_3=0} {p_1,p_2, p_3\ne 0}} \int_{0}^{\infty} d^3 t\,B_{i,j}^w(t_1,t_2,t_3)\,  \exp\Big(\! - \frac{\pi}{\tau_2}\sum_{i=1}^3 t_i |p_i|^2  \Big)\,,
\label{eq:lattint1}
\end{align}
 where $p_i=m_i+n_i\tau$ with $m_i,n_i\in \ZZ$.
The function $B_{i,j}^w(t_1,t_2,t_3)$ is a symmetric function of $(t_1,t_2,t_3)$ which satisfies a Laplace eigenvalue equation\footnote{Here and in subsequent expressions we use the shorthand notation $B(t) \equiv B(t_1,t_2,t_3)$.}
\ie
(\Delta_t - s(s-1)) B_{i,j}^w(t)=0\, ,
\fe
within the domain $t_i>0$, where  $s=3 i + j +1$ with $i,j\in \mathbb{N}$ and $\Delta_t$ is a laplacian expressed in terms of derivatives with respect to $t_i$, as given in \eqref{eq:deltat}. The value of the  index $w\in \NN$ determines the specific combinations of generalised Eisenstein series that arise in $\cE_{i,j}^w$.  Some details of this expression will be described later.  In previous work, such as {\cite{Green:2008bf,DHoker:2018mys}} the function $B_{i,j}^w(t)$ was expressed in the form $V(t)^w  A_{i,j}(\rho(t))$, where $\rho=\rho_1 +i \rho_2$ is a simple complex function of $t_1,t_2,t_3$, as will be reviewed in appendix \ref{sec:geneis}.  
 
The motivation for combining generalised Eisenstein series into expressions such as \eqref{eq:lattint1} can be seen already at $O(N^{-3})$ in \eqref{eq:HN3}. As we will show, the coefficients $\alpha_s,\beta_s,\gamma_s$ (with $s=6,8,10$) of the generalised Eisenstein series at order $1/N^3$ in \eqref{eq:HN3} satisfy a constraint for each value of $s$.  We will further see that this pattern continues to all the orders in $1/N$ that we have evaluated  -- at each order the coefficients are in the basis defined by the integrated lattice sums, $\cE_{i,j}^a$, which is a smaller basis than that defined by  generalised Eisenstein series.   Since the non-holomorphic Eisenstein series in $\cH_N^i$ are already written as lattice sums in \eqref{eq:EisensteinDef}, this shows that the full $\cH_N$ can be written as a lattice sum to all orders in $1/N$, up to at least $O(N^{-7}$) which we have verified explicitly.   

All the explicit results concern properties of  $\cH_N$ at large $N$ and finite $\tau$, and we have not yet found an analytic expression at finite $N$ and $\tau$, as was found for $\cC_N$ in \cite{Dorigoni:2021guq}. However, we will show that the asymptotic large $N$ expansion for both $\cH_N$ and $\cC_N$ can be used to accurately estimate these quantities at finite $N$ for all $\tau$ in the fundamental domain. We do this by combining the exact $N$ expressions for the perturbative terms from \cite{Chester:2019pvm,Chester:2020dja} with the instanton sectors of the $1/N$ expressions in this work truncated to a certain order in $1/N$. The result is shown to numerically match $N=2,3$ expressions for all $\tau$ to good precision, which have been evaluated numerically from the matrix model integrals in \cite{Chester:2021aun}. 

\subsection{Outline}
\label{sec:outline}

The rest of this paper is organised as follows. In Section \ref{sec:largeN} we review the localised expression for $\cH_N$ with emphasis on the zero instanton sector.   We then determine the large-$\lambda$ expansion (where $\lambda=g_{_{YM}}^2 N$ is the 't Hooft coupling) of the correlator, explicitly displaying the first few terms in the $1/\lambda$ expansion at order $N^2$ and $N^0$.  

In Section \ref{sec:half} we will use the above methods to compute the coefficients of half-integer powers of $1/N$ that arise in   $\cH^h_N$.  We will see that this expression satisfies a recursion relation, which generates $\cH^h_N$ to all orders in $1/N$ and at finite $\tau$, thereby  enabling us to determine the expression for $\cH_N^h$ to all orders.   The structure of this expression is remarkably similar to that of $\cC_N$. 

In Section \ref{sec:integer} we consider terms with integer powers of $1/N$  in the large-$N$ expansion of $\cH^i_N$.  {We  will see that the coefficients of each term in the $1/N$ expansion is a  particular combination of generalised Eisenstein series in which contributions from holomorphic cusp forms cancel.} We further show that $\cH^i_N$ admits a four-dimensional lattice sum representation at each order in the large-$N$ expansion.   

 In Section \ref{sec:finiteN} we show how to use the large-$N$ expansion of $\cH_N$ and $\cC_N$ to accurately estimate the value of the integrated correlator at finite values of $N$ and generic values of $\tau$.

 In Section \ref{sec:correlator}, we show how our new results for the $\tau$-independent terms in $\cH_N$ can be used to constrain higher loop corrections to the stress tensor correlator at loop level. 
 
We conclude in Section \ref{sec:disc}  with a discussion of how these results might suggest further insights.  Technical details are relegated to the appendices, and we also include various results in an attached \texttt{Mathematica} file. 

\section{Free energy at large $N$}
\label{sec:largeN}

In this section, we will review the form of the $S^4$ partition function $Z_N(m, \tau)$ of $\cN=2^*$ SYM, which reduces to the partition function of  $\cN=4$ SYM with gauge group $SU(N)$ in the $m\to 0$ limit. We will be particularly interested in computing the large-$N$ expansion of $\mathcal{H}_N$. The emphasis in this section is on  the large-$N$ and large-$\lambda$ expansion, which is not sensitive to instantons. In subsequent sections we will then combine this new data as well as instanton results obtained following \cite{Chester:2020vyz} to fix the large-$N$ finite-$\tau$ expansion to very high orders.  

The  $\cN=2^*$ SYM $S^4$ partition function can be determined using supersymmetric localisation and can be expressed as a matrix model integral \cite{Pestun:2007rz}
 \es{ZFull}{
  Z_N(m, \tau\btau) \! =\! \int d^{N-1} a \,\big( \prod_{i < j}a_{ij}^2\big) e^{- \frac{8 \pi^2}{g_\text{YM}^2} \sum_i a_i^2}  \frac{  \prod_{i\neq j}H(a_{ij})}{ H(m)^{N-1}\! \prod_{i\neq j}H(a_{ij}+ m)}\abs{Z_\text{inst}(m, \tau, a_{ij})}^2 \,,
 }
where $a_{ij}\equiv a_i-a_j$.  The integration is over $N$ real variables $a_i$, $i = 1, \ldots, N$, subject to the constraint $\sum_i a_i = 0$. The function $H(z)=e^{-(1+\gamma)z^2}G(1+iz)G(1-iz)$ is the product of two Barnes G-functions, and $Z_\text{inst}$ is the contribution from instantons localised at the poles of $S^4$.  We can write the instanton partition function as
 \es{ZInstSum}{
  Z_\text{inst}(m, \tau,  a_{ij}) = \sum_{k=0}^\infty q^k  Z_\text{inst}^{(k)} (m, a_{ij}) \,, \qquad {\rm with} \qquad q=e^{2 \pi i \tau} \, , 
 } 
where $Z_\text{inst}^{(k)} (m, a_{ij})$ represents the contribution of the $k$-instanton sector and is normalised such that  $Z_\text{inst}^{(0)} (m,  a_{ij}) = 1$. Explicit expressions for $Z_\text{inst}^{(k)} (m, a_{ij})$ can be found in \cite{Nekrasov:2002qd,Nekrasov:2003rj}. We can then take mass derivatives to get the perturbative contribution
\es{4m}{
 \partial_m^4\log Z_N\big\vert_{m=0}^{pert} =-12\zeta(3)+\sum_{i, j}\langle K'''(a_{ij})\rangle+3\sum_{i,j,k,l}\left[\langle K'(a_{ij})K'(a_{kl})\rangle-\langle K'(a_{ij})\rangle\langle K'(a_{kl})\rangle\right] 
 \,,
 }
where $K(z)\equiv -\frac{H’(z)}{H(z)}$ and expectation values are taken with the Gaussian matrix model measure so that 
\es{ZSUN}{
  \langle \cO(a_{ij}) \rangle =  \frac{1}{\mathcal{N}} \int d^{N} a\,\delta\big(\sum_i a_i\big) \,\big(  \prod_{i < j}a_{ij}^2 \big)e^{-\frac{8 \pi^2 N }{\lambda} \sum_i a_i^2}  \,  \cO(a_{ij})\,,
 } 
 where the normalisation factor $\mathcal{N}$ is chosen such that $\langle 1 \rangle =1$.
 
The instanton terms  in \eqref{ZFull} are contained in Nekrasov partition function, $Z_\text{inst}(m, \tau, a_{ij})$, whose small-$m$ expansion has been studied in \cite{Chester:2019jas, Chester:2020vyz}. In \cite{Chester:2020vyz}, it was already shown how to compute these instanton terms to any order in $1/N$, so in the following we will focus on the perturbative terms \eqref{4m} which we denote by $\partial_m^4\log Z_N\vert_{m=0}^{pert}$.\footnote{Note that the perturbative terms have $k=0$ (zero instanton number), but there is also an infinite set of $k=0$ instanton/anti-instanton terms, which do not contribute to the perturbative sector.}
 
 To evaluate $\partial_m^4\log Z_N\vert_{m=0}^{pert}$, we write the function $K'(z)$ in terms of the integral representation \cite{Russo:2013kea}
  \es{KFourier}{
 K'(z)=-\int_0^\infty d\omega\frac{2\omega[\cos(2\omega z)-1]}{\sinh^2\omega}\,,
 }
 so that the perturbative sector in \eqref{4m} can be written as 
    \es{4mApp}{
\partial_m^4\log Z_N\big\vert_{m=0}^{pert}=&-12\zeta(3)+\int_0^\infty d\omega\frac{8\omega^3 \mathcal{I}(\omega)}{\sinh^2\omega}+\int_0^\infty d\oone d\otwo \frac{12\oone\otwo \mathcal{J}(\oone,\otwo)}{\sinh^2\oone \sinh^2\otwo}\,,
 }
 where we define the 2-body and 4-body expectation values
\es{exp1}{
\mathcal{I}(\omega)\equiv \sum_{i, j}\langle e^{2i\omega a_{ij}}\rangle\,,\qquad \mathcal{J}(\oone,\otwo)\equiv \sum_{i,j,k,l}\left[\langle e^{2i\oone a_{ij}}e^{2i\otwo a_{kl}}\rangle-\langle e^{2i\oone a_{ij}}\rangle\langle e^{2i\otwo a_{kl}}\rangle\right]\,.
}
In \cite{Chester:2019pvm,Chester:2020dja}, it was shown how these expectation values can be computed to any order at large $N$ and finite $\lambda$ using topological recursion, where the expansion takes the form
\es{largeNIJ}{
\mathcal{I}(\omega) &= N^2 \mathcal{I}_{2}(\omega)+N^0 \mathcal{I}_{0}(\omega)+N^{-2} \mathcal{I}_{-2}(\omega)+\dots\,,\\
\mathcal{J}(\oone,\otwo) &= N^2 \mathcal{J}_{2}(\oone,\otwo)+N^0 \mathcal{J}_{0}(\oone,\otwo)+N^{-2} \mathcal{J}_{-2}(\oone,\otwo)+\dots\,.\\
}
The $1/N^2$ coefficients $\mathcal{I}_{i}(\omega)$ are expressed as products of two Bessel functions with the same argument.  For instance, the first two terms are
\es{Is}{
\mathcal{I}_2(\omega)=
\frac{4 \pi ^2  J_1(\frac{ \sqrt{\lambda }\omega}{\pi
   }){}^2}{\lambda \omega^2 }\,,\qquad \mathcal{I}_0(\omega)=
\frac{\sqrt{\lambda}\omega J_1(\frac{ \sqrt{\lambda }\omega}{\pi
   }){}J_2(\frac{ \sqrt{\lambda }\omega}{\pi
   }){}}{12\pi}\,,
}
while higher terms are given in \cite{Chester:2019pvm}. The coefficients $\mathcal{J}_{i}(\oone,\otwo)$ can be expressed as products of four Bessel functions.  For instance,
\begin{align}  \label{Js}
\mathcal{J}_2(\oone,\otwo)&=
\frac{8 \pi  
   J_1(\frac{\sqrt{\lambda } \oone }{\pi })
   J_1(\frac{ \sqrt{\lambda }\otwo}{\pi })}{\sqrt{\lambda }
   (\otwo^2-\oone^2 ) }\left[\textstyle\oone  J_0\left(\frac{\sqrt{\lambda } \oone }{\pi }\right)
   J_1\left(\frac{ \sqrt{\lambda }\otwo}{\pi }\right)-\otwo
   J_1\left(\frac{\sqrt{\lambda } \oone }{\pi }\right)
   J_0\left(\frac{ \sqrt{\lambda }\otwo}{\pi }\right)\right]
   \,, \nn \\ 
   \mathcal{J}_0(\oone,\otwo)&= \mathcal{J}^\text{non-fac}_0(\oone,\otwo)+ \mathcal{J}^\text{fac}_0(\oone,\otwo)\,,
\end{align}
  where we defined the auxiliary functions
  \begin{align}
   \mathcal{J}^\text{non-fac}_0(\oone,\otwo) &\equiv   \frac{\lambda  
   }{12\pi^2
   (\otwo^2{-}\oone^2 ) }\! \left[
 \oone  \textstyle  J_0\left(\frac{\sqrt{\lambda } \oone }{\pi }\right) \textstyle  J_1\left(\frac{\sqrt{\lambda } \otwo }{\pi }\right)
 - \otwo  \textstyle  J_1\left(\frac{\sqrt{\lambda } \oone }{\pi }\right) \textstyle  J_0\left(\frac{\sqrt{\lambda } \otwo }{\pi }\right)
 \right]\\
 &\nn\qquad\qquad\qquad\qquad\times\left[
 \otwo^3  \textstyle  J_1\left(\frac{\sqrt{\lambda } \oone }{\pi }\right) \textstyle  J_2\left(\frac{\sqrt{\lambda } \otwo }{\pi }\right)
 + \oone^3  \textstyle  J_2\left(\frac{\sqrt{\lambda } \oone }{\pi }\right) \textstyle  J_1\left(\frac{\sqrt{\lambda } \otwo }{\pi }\right)
 \right]\\
&\nn+  \frac{\lambda \oone\otwo(\oone^2+\otwo^2)
   }{2\pi^2
   (\otwo^2{-}\oone^2 )^2 } \left[
  \oone \textstyle  J_0\left(\frac{\sqrt{\lambda } \oone }{\pi }\right) \textstyle  J_1\left(\frac{\sqrt{\lambda } \otwo }{\pi }\right)
 - \otwo  \textstyle  J_1\left(\frac{\sqrt{\lambda } \oone }{\pi }\right) \textstyle  J_0\left(\frac{\sqrt{\lambda } \otwo }{\pi }\right)
 \right]\\
&\qquad\qquad\qquad\qquad\times \left[
  \otwo \textstyle  J_0\left(\frac{\sqrt{\lambda } \oone }{\pi }\right) \textstyle  J_1\left(\frac{\sqrt{\lambda } \otwo }{\pi }\right)
 - \oone  \textstyle  J_1\left(\frac{\sqrt{\lambda } \oone }{\pi }\right) \textstyle  J_0\left(\frac{\sqrt{\lambda } \otwo }{\pi }\right)
 \right]\,, \nn 
 \end{align}
  \begin{align}
   \mathcal{J}^\text{fac}_0(\oone,\otwo)& \equiv-\frac{7 \lambda  \otwo^2 J_1\left(\frac{\sqrt{\lambda } \oone }{\pi }\right){}^2
   J_1\left(\frac{\sqrt{\lambda }\otwo }{\pi }\right){}^2}{12 \pi ^2}-\frac{\lambda  \otwo^2 
   J_1\left(\frac{\sqrt{\lambda } \oone }{\pi }\right){}^2 J_0\left(\frac{\sqrt{\lambda }\otwo
   }{\pi }\right){}^2}{2 \pi ^2}\\
   &\nn-\frac{7 \lambda  \oone ^2
   J_1\left(\frac{\sqrt{\lambda } \oone }{\pi }\right){}^2 J_1\left(\frac{\sqrt{\lambda }\otwo
  }{\pi }\right){}^2}{12 \pi ^2}+\frac{\lambda   \oone \otwo
   J_0\left(\frac{\sqrt{\lambda } \oone }{\pi }\right) J_1\left(\frac{\sqrt{\lambda }
   \oone }{\pi }\right) J_0\left(\frac{\sqrt{\lambda }\otwo }{\pi }\right) J_1\left(\frac{\sqrt{\lambda }\otwo
   }{\pi }\right)}{2 \pi ^2}\\
   &+\frac{\lambda   \oone \otwo
   J_1\left(\frac{\sqrt{\lambda } \oone }{\pi }\right) J_2\left(\frac{\sqrt{\lambda }
   \oone }{\pi }\right) J_2\left(\frac{ \sqrt{\lambda }\otwo}{\pi }\right) J_1\left(\frac{
   \sqrt{\lambda }\otwo}{\pi }\right)}{12 \pi ^2}-\frac{\lambda  \oone ^2
   J_0\left(\frac{\sqrt{\lambda } \oone }{\pi }\right){}^2 J_1\left(\frac{
   \sqrt{\lambda }\otwo}{\pi }\right){}^2}{2 \pi ^2} \nn
   \,.
\end{align}
Note that at order $O(N^0)$ we have separated the terms into $   \mathcal{J}^\text{fac}_0$, which factorises in terms of $\otwo$ and $\oone$, and $   \mathcal{J}^\text{fac}_0$, which does not factorise. Higher order terms take a similar form; they can be found in \cite{Chester:2020dja} and up to order $O(N^{-8})$ in the attached \texttt{Mathematica} file. 

The large-$\lambda$ expansion of the 2-body term $\mathcal{I}(\omega)$ in \eqref{4mApp} is straightforward at all orders in $1/N^2$. As shown in \cite{Binder:2019jwn}, we start by writing bilinears of Bessel functions in the Mellin-Barnes form
\es{mbbessel}{
J_\mu(x)J_\nu(x) =\oint  \frac{ds}{2\pi i}\frac{\Gamma(-s)\Gamma(2s+\mu+\nu+1)\left(\frac x2\right)^{\mu+\nu+2s}}{\Gamma(s+\mu+1)\Gamma(s+\nu+1)\Gamma(s+\mu+\nu+1)} \,,
}
where the contour separates the poles of each Gamma function. We then perform the $\omega$ integrals in \eqref{4mApp} using the identity
\es{id}{
\int_0^\infty d\omega\ \frac{\omega^{a}}{\sinh^2\omega} = \frac{1}{2^{a-1}}\Gamma(a+1)\zeta(a) \,,
}
 and then close the contour to the left to get an expansion in $1/\lambda$. For instance, using the explicit expressions in \eqref{Is} we get
 \es{Ieasy}{
 &\int_0^\infty d\omega\frac{8\omega^3 \mathcal{I}(\omega)}{\sinh^2\omega}=N^2\Big[\frac{16 \pi ^2}{\lambda
   }-\frac{32 \pi ^2}{\lambda ^{\frac{3}{2}}}+\frac{24 \pi ^2 \zeta (3)}{\lambda ^{\frac{5}{2}}}+O(\lambda^{-\frac72})\Big]\\
   &\qquad\qquad\qquad\qquad+N^0\Big[\frac{4 \pi ^2
   \sqrt{\lambda }}{15}-\frac{13 \pi ^2}{16 \lambda ^{\frac{3}{2}}}-\frac{75 \pi ^2 \zeta (3)}{32 \lambda
   ^{\frac{5}{2}}}+O(\lambda^{-\frac72})\Big]+O(N^{-2})\,,\\
 }
 where the expansion at each order in $1/N^2$ including a finite number of positive powers of $\lambda$, plus an infinite number of negative powers.
 
 The large-$\lambda$ expansion of the 4-body terms $\mathcal{J}(\oone,\otwo)$ in \eqref{4mApp} is much harder {to evaluate} because the dependence on $\oone$ and $\otwo$ does not in general factorise. As a result, even if we write each pair of Bessel functions in Mellin-Barnes form, it is still difficult to perform the $\oone,\otwo$ integrals. In \cite{Chester:2020dja}, the 4-body terms were instead computed at large $\lambda$ by numerically evaluating the $\oone,\otwo$ integrals in \eqref{4mApp} for many values of finite $\lambda$, and then using a numerical fit to guess the large $\lambda$ values. This method only worked for the lowest few orders in $1/\lambda$ at each order in $1/N^2$, where it gave 
\begin{align}
 \label{secondTermNFac}
\int_0^\infty  & d\oone  d\otwo\frac{12\oone\otwo \mathcal{J}(\oone,\otwo)}{\sinh^2\oone\sinh^2\otwo}= \nn \\
&N^2\Big[6+\frac{96 \zeta (3)}{\lambda ^{\frac{3}{2}}}-\frac{288 \zeta (5)}{\lambda
   ^{\frac{5}{2}}}-\frac{144 \zeta (3)^2}{\lambda
   ^{3}}-\Big(\frac{16 \pi ^2}{\lambda
   }-\frac{32 \pi ^2}{\lambda ^{\frac{3}{2}}}+\frac{24 \pi ^2 \zeta (3)}{\lambda ^{\frac{5}{2}}}\Big)+O(\lambda^{-\frac72})\Big] \nn \\
&+N^0\Big[\Big(2-\frac{4 \pi ^2}{15}\Big) \lambda^\half +O(\lambda^{0})\Big]+N^{-2}\Big[\Big(\frac{\pi ^2}{504}-\frac{1}{60}\Big)\lambda^{\frac32}+O(\lambda)\Big]  \\
      &+N^{-4}\Big[-\frac{\lambda^3}{120960}+O(\lambda^{\frac52})\Big]+O(N^{-6})\, . \nn
\end{align}
Note that the $\pi^2$ terms here exactly cancel those from the 2-body term in \eqref{Ieasy} up to the order shown.
 
We now show how to evaluate the 4-body terms at large $\lambda$ analytically to any order. Let us begin with the leading $N^2$ term $\mathcal{J}_2(\oone,\otwo)$ in \eqref{Js}:
    \begin{align} \label{tree221}
 I^{(0)}(\lambda)& \equiv \int_0^\infty \! d\oone d\otwo\frac{12\oone \otwo \mathcal{J}_2(\oone,\otwo)}{\sinh^2\oone\sinh^2\otwo}\\
 &=\int_0^\infty\!  d\oone d\otwo \frac{96 \pi  \oone\otwo 
   J_1 \! \left(\frac{\sqrt{\lambda } \oone }{\pi }\right)\!
   J_1\! \left(\frac{ \sqrt{\lambda } \otwo}{\pi }\right)}{\sqrt{\lambda }
   (\otwo^2 {-} \oone^2 ) \sinh^2\oone\sinh^2\otwo}\left[\textstyle\oone  J_0\! \left(\frac{\sqrt{\lambda } \oone }{\pi }\right)\!
   J_1\! \left(\frac{\sqrt{\lambda }\otwo }{\pi }\right) \! -\otwo
   J_1\! \left(\frac{\sqrt{\lambda } \oone }{\pi }\right)\!
   J_0\! \left(\frac{\sqrt{\lambda }\otwo }{\pi }\right)\right]\,. \nonumber
  \end{align}
We start by applying the Bessel kernel identity
 \es{BesKern}{
\frac{  \oone  J_0\left(\frac{\sqrt{\lambda } \oone }{\pi }\right)
   J_1\left(\frac{\sqrt{\lambda }\otwo }{\pi }\right)-\otwo
   J_1\left(\frac{\sqrt{\lambda } \oone }{\pi }\right)
   J_0\left(\frac{ \sqrt{\lambda }\otwo}{\pi }\right)}{\otwo^2 - \oone^2}
    = \sum_{\ell=1}^\infty \frac{4  \ell \pi  J_{2\ell} \left(\frac{\sqrt{\lambda } \oone }{\pi }\right) J_{2\ell} \left(\frac{\sqrt{\lambda } \otwo }{\pi }\right)}{ \oone\otwo \sqrt{\lambda}} \,,
 } 
to write \eqref{tree221} as an infinite sum of terms
 \es{IAgain}{
  I^{(0)}(\lambda) =  \sum_{\ell=1}^\infty \frac{384 \pi^2 \ell }{\lambda}    \int_0^\infty  d\oone d\otwo \frac{  
   J_1\left(\frac{\sqrt{\lambda } \oone }{\pi }\right) J_{2\ell} \left(\frac{\sqrt{\lambda } \oone }{\pi }\right)
  }{
   \sinh^2\oone}\frac{  
   J_1\left(\frac{\sqrt{\lambda } \otwo }{\pi }\right) J_{2\ell} \left(\frac{\sqrt{\lambda } \otwo }{\pi }\right)
  }{
   \sinh^2\otwo}  \,.
 }
Since this expression now factorises in $\oone$ and $\otwo$, we can convert each pair of Bessel functions to Mellin-Barnes form using \eqref{mbbessel}, shift the contour integral variables by $1-\ell$, and perform the $\oone$ and $\otwo$ integrals using \eqref{id}, which gives
\es{I3}{
 I^{(0)}(\lambda) =\oint\frac{d\sint}{2\pi i}\frac{d\tint}{2\pi i} p(\sint,\tint) c(\sint,\tint)\,,\qquad c(\sint,\tint) =\sum^{\infty}_{\ell=1}c_\ell(\sint,\tint)\,,
} 
where the $\ell$-independent prefactor is
\es{p}{
p(\sint,\tint) = \frac{3 \lambda^{\sint
   +\tint +2}\sin(\pi\sint)\sin(\pi\tint)\Gamma (2 \sint +4)^2
   \Gamma (2 \tint +4)^2 \zeta (2 \sint +3) \zeta (2 \tint +3)}{  2^{4 \sint +4 \tint +3} \pi ^{2 (\sint +\tint +2)}  }\,,
}
while the $\ell $-dependent terms are 
\es{csig}{
c_\ell(\sint,\tint)=\frac{\ell (\ell-\sint -3) (\ell-\tint -3) \Gamma (\ell-\sint -3)
   \Gamma (\ell-\sint -1) \Gamma (\ell-\tint -3) \Gamma (\ell-\tint
   -1)}{\pi ^2 \Gamma (\ell+\sint +2) \Gamma (\ell+\sint +3)
   \Gamma (\ell+\tint +2) \Gamma (\ell+\tint +3)}\,.
}
We can now perform the contour integrals by closing the contours to the left, which includes poles where $\sint$ and $\tint$ are each negative numbers, as well as poles where the sum $\sint+\tint$ is a negative number. The details of this pole counting are summarised in appendix \ref{app:details}. The result matches the leading large-$N$ term in \eqref{secondTermNFac}, but now we can extend the series to any order in $1/\lambda$ systematically. 

The 4-body term at order $N^0$ is more complicated, as $ \mathcal{J}_0(\oone,\otwo)$ in \eqref{Js} now includes twice the non-factorised denominator $\otwo^2-\oone^2$.\footnote{This additional complication is only relevant for the $O(N^0)$ term.} To make the $\oone,\otwo$ factorise, we must therefore apply the Bessel kernel identity \eqref{BesKern} twice, so the resulting expression will be a double infinite sum. In appendix \ref{app:details}, we show how the above method is modified to compute the large-$\lambda$ expansion in this case. After combining these 4-body results with the 2-body expression \eqref{Ieasy}, we find that that the expansion of the full perturbative sector \eqref{4mApp} is given by
\begin{align}  \label{4mApp2}
\partial_m^4  \log & Z_N  \big\vert_{m=0}^{pert} =
N^2\Big[6+\frac{96 \zeta (3)}{\lambda ^{\frac{3}{2}}}
-\frac{288 \zeta
   (5)}{\lambda ^{\frac{5}{2}}}
     -\frac{144 \zeta (3)^2}{\lambda
   ^3}
   -\frac{3375 \zeta (7)}{4 \lambda
   ^{7/2}}
     -\frac{1080 \zeta (3)
   \zeta (5)}{\lambda ^4}
   +O(\lambda^{-\frac92})\Big] \nn \\
   &+\Big[
   2 \sqrt{\lambda } -\frac{96 \zeta (3)+23}{10}
    -\frac{3 \zeta (3)}{\lambda }+\frac{117 \zeta (3)}{16 \lambda
   ^{\frac{3}{2}}}  -\frac{45 \zeta (5)}{8
   \lambda ^2}+O(\lambda^{-\frac52}) \Big]+O(N^{-2})
\,.
\end{align}
We note that the $\pi^2$ terms from the 2-body contribution \eqref{Ieasy} cancel at all orders, and the $\lambda^0 N^0$ term also receives a contribution from the first term in \eqref{4mApp}. All higher $1/N^2$ corrections can be evaluated similarly. In the attached \texttt{Mathematica} file, we show explicit results to higher orders in both $1/N^2$ and $1/\lambda$. They take a similar form to the above expressions, except the number of positive powers of $\lambda$ grows with each order in $1/N^2$, and all subsequent $\lambda^0$ terms are simple fractions. In the following sections we will use these perturbative data, combined with the instanton results obtained following \cite{Chester:2020vyz}, to determine the coefficients of  the large-$N$ finite-$\tau$ expansion.

\section{Half-integer powers of $1/N$ at finite $\tau$}
\label{sec:half}

We now consider the large-$N$ and finite-$\tau$ expansion. As discussed around \eqref{eq:HN}, this expansion can be divided into terms with half-integer powers of $N$ contained in $\cH_N^h(\tau)$, as well as terms with integer power of $N$ in $\cH_N^i(\tau)$.  In this section we will consider the form of  $\cH^h (\tau)$. We can detemine the coefficients of the powers of $1/N$ using the perturbative sector of the localisation expression \eqref{4m}, as well as the instanton terms. As discussed earlier, the instanton terms were obtained  in \cite{Chester:2020vyz}, so we will mainly discuss perturbative terms.

The perturbative sector at large $N$ and finite $\tau$ only depends on $\tau_2$, and can be obtained from the large-$N$ and large-$\lambda$ expansion in \eqref{4mApp2} by simply setting $\lambda={4\pi N}/{ \tau_2}$ and reorganising the large-$N$ expansion. In \cite{Chester:2020vyz} it was observed that $\cH_N^h(\tau)$ is entirely written in terms of non-holomorphic Eisenstein series $E(s;\tau\btau)$, which have the well-known  Fourier-mode decomposition
\es{EisensteinExpansion}{
  E(s; \tau\btau)
   &= \frac{2 \zeta(2 s)}{\pi^s}{\tau_2^s} +   \frac{2\Gamma(s - \frac 12)}{\pi^{s-\frac12}\Gamma(s)} \zeta(2s-1)\tau_1^{1-s} \\
    &{}+ \frac{4 \sqrt{\tau_2}}{\Gamma(s)} \sum_{k\ne 0} \abs{k}^{s-\half}
    \sigma_{1-2s}(k) \, 
      K_{s - \frac 12} (2 \pi  \abs{k}\tau_2) \, e^{2 \pi i k \tau_1} \, ,
 }
 where the divisor sum $\sigma_p(k)$ is defined as $\sigma_p(k)=\sum_{d|k}  d^p$ with $d>0$, and $K_{s - \frac 12}$ is a Bessel function of the second kind. The contribution  to $\cH_N^h(\tau)$ coming from a particular non-holomorphic Eisenstein series, $E(s;\tau)$, {is in fact uniquely fixed by the form of the two $k=0$ terms} in the Fourier series \eqref{EisensteinExpansion}.  We can therefore determine $\cH_N^h(\tau)$ to any order in $1/N$ and finite $\tau$ by simply comparing to the perturbative data from \eqref{4mApp2} (and the higher order terms given in the attached \texttt{Mathematica} notebook). We can then double check that the $k>0$ terms in \eqref{EisensteinExpansion} are consistent with the $k>0$ terms derived in \cite{Chester:2020vyz} (as well as new data obtained following the methods outlined in \cite{Chester:2020vyz}). We find that $\cH_N^h(\tau\btau)$ takes the following form
\es{eq:Hnh}{
\cH_N^h(\tau\btau) = \sum_{j,\ell=0}^\infty  N^{{\half -\ell-2j}} d_\ell^{j}\, E({\scriptstyle\frac32}+\ell; \tau\btau) \,,
}
where the coefficients $d_\ell^{j}$ for the lowest few $j$ are found to be
\allowdisplaybreaks{
\begin{align} \label{ds}
d_\ell^{0}&=-\frac{ (\ell+1) (\ell+3) \Gamma
   \left(\ell-\frac{1}{2}\right) \Gamma
   \left(\ell+\frac{3}{2}\right)^2}{4^{\ell-1}\pi ^{\ell+\frac92} \Gamma (\ell+1)}\,,\\
   d_\ell^{1}&=\frac{ (\ell+1) (2 \ell+13) (\ell (2 \ell+7)+9) \Gamma
   \left(\ell+\frac{1}{2}\right) \Gamma \left(\ell+\frac{3}{2}\right)
   \Gamma \left(\ell+\frac{5}{2}\right)}{3 \cdot 4^{ \ell+2} \pi ^{\ell+\frac{11}{2}}(\ell+2) \Gamma (\ell+1)}
   \,,\cr
   d_\ell^{2}&=-\frac{(\ell+1)^2 (\ell (2 \ell+7)+15) (4 \ell (5 \ell+52)+219)
   \Gamma \left(\ell+\frac{1}{2}\right) \Gamma
   \left(\ell+\frac{3}{2}\right) \Gamma \left(\ell+\frac{11}{2}\right)}{45\cdot 4^{\ell+5} \pi ^{\ell+\frac{13}{2}} 
   \Gamma (\ell+4)}
      \,,\cr
       d_\ell^{3}&=\frac{(\ell+1)^2 (\ell (2 \ell+7)+21) (560 \ell^4 {+} 10752 \ell^3 {+} 59272 \ell^2 {+} 124800 \ell  {+}  75771 ) \Gamma\! \left(\ell+\frac{1}{2}\right)\! \Gamma\!
   \left(\ell+\frac{3}{2}\right)\! \Gamma\!
   \left(\ell+\frac{15}{2}\right)}{2835\cdot 4^{\ell+8} \pi ^{\ell+\frac{15}{2}}  \Gamma (\ell+5)}
            \,. \nonumber
\end{align}}
The integrated correlator $\cC_N(\tau\btau)$ can be written in a similar form
\es{eq:Gnh}{
\cC_N(\tau\btau) =\frac{N^2}{4}+ \sum_{j,\ell=0}^\infty  N^{{\half -\ell-2j}} {\tilde d}_\ell^{j}\, E({\scriptstyle\frac32}+\ell; \tau\btau) \,,
}
where the lowest few coefficients are \cite{Dorigoni:2021guq}
\begin{align} \label{ds2}
\tilde d_\ell^{0}&=\frac{ (\ell+1) \Gamma \left(\ell-\frac{1}{2}\right)
   \Gamma \left(\ell+\frac{3}{2}\right) \Gamma
   \left(\ell+\frac{5}{2}\right)}{ 4^{ \ell+1} \pi ^{\ell+\frac{9}{2}} \Gamma (\ell+1)}
   \,,\\
 \tilde  d_\ell^{1}&=-\frac{(\ell+1)^2 (2 \ell+13) \Gamma
   \left(\ell+\frac{3}{2}\right) \Gamma \left(\ell+\frac{5}{2}\right)^2}{3 \cdot2^{2 \ell+7} \pi ^{\ell+\frac{11}{2}} 
   \Gamma (\ell+3)}
      \,,\cr
   \tilde    d_\ell^{2}&=\frac{ (\ell+1)^2 \left(20 \ell^2+208 \ell+219\right)
   \Gamma \left(\ell+\frac{3}{2}\right) \Gamma
   \left(\ell+\frac{5}{2}\right) \Gamma \left(\ell+\frac{11}{2}\right)}{45\cdot 2^{2 \ell+13} \pi ^{\ell+\frac{13}{2}}
   \Gamma (\ell+4)}
            \, , \cr
        \tilde    d_\ell^{3}&=-\frac{(\ell+1)^2 \left(560 \ell^4+10752 \ell^3+59272
   \ell^2+124800 \ell+75771\right) \Gamma\! \left(\ell+\frac{3}{2}\right)\! \Gamma\!
   \left(\ell+\frac{5}{2}\right)\! \Gamma\!
   \left(\ell+\frac{15}{2}\right)}{2835\cdot 2^{2 \ell+19} \pi ^{\ell+3}  \Gamma (\ell+\frac{15}{2})}
            \,.\nonumber
\end{align}
While the coefficients in both \eqref{ds} and \eqref{ds2} may seem quite complicated, it turns out for all $j$ they are simply related by
\es{dratio}{
\left( 3 +  8\ell +4 \ell^2\right) {d_\ell^j}=-32\left( 3 + 6 j +  7\ell + 2 \ell^2\right)   {\tilde d_\ell^j}\,.
}
These simple polynomials in $\ell,j$ can be interpreted as coming from derivatives {with respect to}  $\tau$ and $N$ acting on \eqref{eq:Hnh} and \eqref{eq:Gnh}, which implies the formal relation
\es{dratio2}{
\Delta_\tau \cH_N^h(\tau)=-8\left( 3 -3\partial_N+2\Delta_\tau\right)  \cC_N(\tau\btau)+6N^2-12N\,,
}
where the polynomial in $N$  cancels the $N^2/4$ term in $\cC_N(\tau\btau)$. We can eliminate $\partial_N$, whose interpretation is confusing for integer $N$, by acting with $\partial_N$ on the recursion relation for $\cC_N(\tau\btau)$ in \eqref{eq:Grec} and combining with shifted versions of \eqref{dratio2}, which yields the recursion relation given in \eqref{eq:Hrec}. This recursion relation, or equivalently \eqref{dratio}, can be used to generate $\cH_N^h(\tau)$ to any desired order in $1/N$ from the known $\cC_N(\tau\btau)$ \cite{Dorigoni:2021guq}.  The new recursion relation essentially puts $\cH^h_N(\tau\btau)$ on the same level as $C_N(\tau\btau)$, which is known to any order in $1/N$. However, the full second integrated correlator $\cH_N(\tau\btau)$ does not obey the recursion relation \eqref{eq:Hrec}. In particular, the integer powers in $1/N$ expansion $\cH^i_N(\tau\btau)$ behave differently, as we will discuss in the next section.
  
  \section{Integer powers of $1/N$ at finite $\tau$}
\label{sec:integer}

We will now discuss {$\cH_N^i(\tau)$, which contains the  integer powers of $1/N$ in the large-$N$  expansion.} The terms up to   $1/N^3$ were  obtained in   \cite{Chester:2020vyz} and are reproduced in \eqref{eq:HN3}.   The coefficients of these terms are sums of  generalised Eisenstein series $\cE(s,s_1,s_2;\tau)$ with $s_1, s_2 \in \NN + \half$, as well as $\tau$-independent pure numbers.   {We find that the} generalisation of \eqref{eq:HN3}  to arbitrary orders in $1/N$ takes the following form,   \es{eq:Hni}{
\cH_N^i(\tau) =\sum_{j=0}^\infty\frac{C_j}{N^{2j}} +\sum_{j=0}^\infty\sum_{\ell=0}^\infty\sum_{r=0}^{\floor{\ell/2}} \sum_{s=4+\ell,6+\ell,\dots}^{3\ell +6j+4} \frac{\alpha_{j,\ell,r,s}}{ N^{{1 +\ell+2j}} } \cE(s,{\scriptstyle \frac32}+r,{\scriptstyle \frac32}+\ell-r;\tau) \, .
}
The $\tau$-independent terms $C_j$ are associated with supergravity contributions, and  terms containing $\cE(s,s_1,s_2;\tau)$  are due to the stringy effects. The choice of the eigenvalues $s$ is to ensure that the terms at order $1/N^{n}$ obeys a boundary condition such that the corresponding  (integrated) $d^{2(2n+1)}R^4$ vertices in the holographic string amplitudes contribute only up to genus-$(2n+1)$, which was previously observed in \cite{Chester:2020vyz}. Whereas the choice of the sources, {which are labelled by $s_1, s_2$}, can be understood as a consequence of the relation \eqref{eq:Hrec2}.

The $\tau$-independent terms $C_j$ {in \eqref{eq:HN3}} were not computed in \cite{Chester:2020vyz}, because they can only be fixed from the perturbative sector. Using our new perturbative data, {we find that 
\es{Cs}{
C_0=-\frac{ 96 \zeta (3)+23}{10}\,,
}
 while for $j>0$ they all appear to be  rational numbers. For instance, $C_1=-\frac{1}{10}\,,$ and $C_2=\frac{1}{14}$. It  is straightforward (but laborious) to compute higher orders from the perturbative results.}

As for the $\cE(s,s_1,s_2;\tau)$ terms in $\cH_N^i(\tau)$, we first note that the $\cE(s,s_1,s_2;\tau)$ have the Fourier expansion
\es{genEsketch}{
\cE(s,s_1,s_2;\tau) = \sum_{n,m= 0 }^\infty \cE^{(n,m)}(s,s_1,s_2; \tau_2)q^n \bar{q}^m\, .
}
The coefficients  $\cE^{(n,m)}(s,s_1,s_2;\tau_2)$ correspond to terms in the $n$-instanton/$m$-anti-instanton sector with total instanton number $k=n-m$.  Further discussion of $\cE^{(n,m)}(s,s_1,s_2; \tau_2)$ can be found in appendix \ref{sec:genais}.  
The coefficients of terms in the $1/N$ expansion of  $\cH_N^i(\tau)$ can in principle be fixed directly by comparing with explicit results obtained from localisation for certain choices of $(m,n)$.  This was the procedure used in  \cite{Chester:2020vyz} where all the terms up to $O(1/N^3)$ were determined. This however becomes more and more challenging as we consider higher-order terms. In order to simplify the computation, we will assume that   $\cH_N^i(\tau)$ satisfies certain conditions at all orders in $1/N$  that are satisfied by low-order terms.  We will then check that the resulting expression indeed reproduces the localisation result.  The conditions that will be imposed are the following:
\begin{itemize}
 \item   Up to order $1/N^3$ we  observed in \eqref{eq:Hrec2} that the action of  $\Delta_\tau$ on $\cH_N^i(\tau)$  results in terms bilinear in Eisenstein series that are proportional to those in the square of $\cC_N(\tau\btau)$. Assuming that this continues to hold at higher orders in $1/N$ provides strong constraints on  the coefficients of the generalised Eisenstein series that arise in the $1/N$ expansion.

 \item The second condition is that the coefficients $\alpha_{j,\ell,r,s}$ of the expansion \eqref{eq:Hni} in generalised Eisenstein series are related for each eigenvalue $s$ in such a manner that contributions coming from holomorphic cusp forms cancel. As will be discussed in the next subsection individual $\cE(s,s_1,s_2;\tau)$ with $s\ge 6$ generically have such cusp forms, but they are not present in the localisation result of $\cH_N$.  Such cancellation was present (but not appreciated) in the order $1/N^3$ expression in  \cite{Chester:2020vyz}. Here it will be extended to arbitrary order in $1/N$.   Furthermore, we will  show that the combination of $\cE(s,s_1,s_2;\tau)$ for which the cusp forms cancel is one for which there is lattice sum representation. 
 
\end{itemize}

These supplementary conditions  significantly reduce the number of independent coefficients $\alpha_{j,\ell,r,s}$ at each order in $1/N$. For example, at order  $1/N^3$, although there are apparently $10$ coefficients only $4$ of these are independent once the above extra conditions are satisfied.   At order $1/N^4$ there appear to be $13$ independent coefficients, but this is reduced to $7$ after imposing the above conditions, and at order $1/N^5$ the $34$ apparently independent coefficients are reduced to $16$. Having constrained the coefficients in the $1/N$ expansion  in this manner, we will then use new perturbative and non-perturbative data obtained in this paper from the localisation computation to fix these coefficients. In this way, we have determined $\cH_N^i(\tau)$ completely up to order $O(N^{-7})$. It is important to stress that we have checked that the results agree with  much more localisation data than is needed to fix the coefficients, This procedure therefore provides non-trivial consistency checks on the properties observed in the  lower-order results. 

Some examples of the coefficients of powers of $1/N$ will be  displayed later in this section, but in general the explicit expressions at higher orders in $1/N$  are rather lengthy, so we relegate them to the attached \texttt{Mathematica}  file.

\subsection{Holomorphic cusp forms and their cancellation}
\label{sec:cusp}

In \cite{Chester:2020vyz}, then generalised in \cite{Fedosova:2022zrb}, it was shown how one can construct particular solutions, $\cE_{{\rm p}}(s,s_1,s_2;\tau)$, to the Laplace equation \eqref{eq:laplacewsource}. However, in the earlier works \cite{Dorigoni:2021jfr,Dorigoni:2021ngn} it was demonstrated that when $s_1,s_2\in \NN$ (relevant for modular graph functions), such particular solutions do not always correspond to modular invariant functions.  In such cases it  was shown that the failure of modularity of $\cE_{{\rm p}}(s,s_1,s_2;\tau)$ requires the addition of a suitable  homogeneous solution.

Using the results of \cite{FKRinprogress}, a similar argument can be given for the case $s_1,s_2 \in \mathbb{N}+\frac{1}{2}$ that is of present interest.
It can be shown that the generalised Eisenstein series $ \cE$ differs from the particular solution $\cE_{{\rm p}}$ constructed via \cite{Chester:2020vyz,Fedosova:2022zrb} in the purely instantonic or anti-instantonic sector, i.e. for $(n,m)=(n,0)$ or $(n,m)=(0,m)$, and takes the form 
\begin{equation}\label{eq:Einst}
 \cE^{(n,0)}(s,s_1,s_2;\tau_2)=  \cE^{(n,0)}_{{\rm p}}(s,s_1,s_2;\tau_2) + \sum_{\Delta\in \cS_{2s}} \lambda_{\Delta}(s,s_1,s_2) \frac{a_\Delta(n)}{n^s} \sqrt{n\tau_2 } e^{2\pi  n \tau_2} K_{s-\frac{1}{2}}(2 \pi n \tau_2)\,.
\end{equation}
The asymptotic expansion at the cusp of $\cE^{(n,0)}_{{\rm p}}(s,s_1,s_2;\tau_2)$ produces a perturbative expansion in $\tau_2^{-1}$, which are the terms that we matched to the instanton sector of $\cH_N^i(\tau)$ here and in \cite{Chester:2020vyz}.  The second term in \eqref{eq:Einst} was missed in the earlier analysis of \cite{Chester:2020vyz, Fedosova:2022zrb}\footnote{As already mentioned, and will be shown in detail later, the holomorphic cusp forms cancel out in the final expression of the integrated correlator, therefore the final result presented in \cite{Chester:2020vyz} is not affected by this. } and it involves a sum over the vector space $\cS_{2s}$ of  all holomorphic cusp forms with modular weight $2s$, i.e. we need to sum over all $\Delta \in \cS_{2s}$ of the form
\begin{equation}
\Delta(\tau) \equiv \sum_{n=1}^\infty a_\Delta(n) q^n \in \cS_{2s}\,,
\end{equation}
with Hecke normalisation $a_\Delta(1)= 1$, and again $q=\exp(2\pi i \tau)$.

{Following \cite{Dorigoni:2021jfr,Dorigoni:2021ngn}}  {we consider the contribution of a  fixed cusp form, $\Delta \in \cS_{2s}$, to $\cE(s,s_1,s_2;\tau)$}  {coming from the second term in \eqref{eq:Einst} when summed over all instanton and all anti-instanton sectors, which takes the form
\begin{equation}
H_{\Delta}(\tau)\equiv \sum_{n=1}^\infty  \frac{a_\Delta(n)}{n^s} \sqrt{n\tau_2 }  K_{s-\frac{1}{2}}(2 \pi n \tau_2)\big( e^{2\pi i n \tau_1}+e^{-2\pi i n \tau_1}\big)\, .
\end{equation}
This function is not modular invariant but it does provide for a homogeneous solution to the Laplace equation \eqref{eq:laplacewsource}. We refer to the combination $ \lambda_{\Delta}(s,s_1,s_2)H_{\Delta}(\tau)$ as the \textit{cusp form $\Delta$ contribution} to $\cE(s,s_1,s_2;\tau)$, see appendix \ref{sec:genais} for more details.}

For the case $s_1,s_2\in \NN$ relevant for modular graph functions, the coefficient $\lambda_\Delta(s,s_1,s_2)$ was determined in \cite{Dorigoni:2021jfr,Dorigoni:2021ngn}, while for $s_1,s_2 \in \mathbb{N}+\frac{1}{2}$ we use the results of \cite{FKRinprogress} and find that $\lambda_\Delta(s,s_1,s_2)$ takes the schematic form
\begin{equation}
\lambda_{\Delta} (s,s_1,s_2)= q(s,s_1,s_2) (-1)^{s_1}\frac{\Lambda(\Delta; s+1-s_1-s_2)\Lambda(\Delta; s+s_1-s_2)}{\pi \langle \Delta,\Delta\rangle}\,,
\label{eq:lvalue}
\end{equation}
where $q(s,s_1,s_2)$ is a rational number depending on $s, s_1, s_2$, and symmetric under the exchange $s_1\leftrightarrow s_2$. We denote by $\Lambda(\Delta;t)$ the completed $ {\rm L}$-value of $\Delta$, 
\begin{equation}
\Lambda(\Delta; t) \equiv (2\pi)^{-t} \Gamma(t) {\rm L}(\Delta ; t)\, ,
\end{equation}
which is defined for $t\in \mathbb{C}$ via analytic continuation of the Dirichlet series
\begin{equation}
 {\rm L}(\Delta; t) \equiv \sum_{n=1}^\infty \frac{a_\Delta(n)}{n^t}\, ,
\end{equation}
and the denominator in \eqref{eq:lvalue}, $\langle \Delta,\Delta\rangle$, is the Petersson norm of $\Delta\in \cS_{2s}$.\footnote{The precise definition of the Petersson norm is not essential for the following discussion.} Although not manifest from \eqref{eq:lvalue}, for $\Delta\in \cS_{2s}$ the coefficients $\lambda_\Delta$ are symmetric under the exchange $s_1\leftrightarrow s_2$ as a consequence of the functional equation 
\begin{equation}
\Lambda(\Delta,t) = (-1)^s \Lambda(\Delta,2s-t)\,.
\end{equation}

Therefore, whenever the eigenvalue $s$ is such that $\dim \mathcal{S}_{2s}\neq 0$, the particular solution to the inhomogeneous Laplace equation \eqref{eq:laplacewsource}, as constructed from the particular solution $\cE^{(n,0)}_{{\rm p}}(s,s_1,s_2; \tau_2)$ \cite{Chester:2020vyz,Fedosova:2022zrb}, in general does not lead to a modular invariant solution. Only by adding the second term in \eqref{eq:Einst}, do we then arrive at the modular invariant solution to \eqref{eq:laplacewsource} that we denote by $\cE(s,s_1,s_2;\tau)$. 

The dimension of the vector space of holomorphic cusp forms $\mathcal{S}_{2s}$ is
\begin{equation}\label{eq:dimS}
 \dim \mathcal{S}_{2s} = \left\lbrace \begin{array}{lc}
\left\lfloor \frac{2s}{12}\right\rfloor -1 & 2s\equiv 2\, {\rm mod }\,12\,.\\[2mm]
 \left\lfloor \frac{2s}{12}\right\rfloor \phantom{-1}& \mbox{otherwise}
\end{array}\right.
\end{equation}
Since $\dim \mathcal{S}_{12}=1$, we see from our ansatz \eqref{eq:Hni} that cusp forms become relevant starting at order $O(N^{-3})$ which is the first instance where the eigenvalue $s=6$ appears. In this case, the unique Hecke normalised cusp form of weight $12$ is given by the Ramanujan cusp form $\Delta_{12} = \sum_{n=1}^\infty\tau(n) q^n$ where $\tau(n)$ (not to be confused with the coupling $\tau$) denotes the Ramanujan tau function. 

Let us consider the $s=6$ terms at order $O(N^{-3})$ as an example, which were denoted in \eqref{eq:HN3} as
\es{N3}{
\cH_{N}^i(\tau)\big\vert^{s=6}_{ N^{-3}}=\frac{1}{N^3}\Big[\alpha_6\,\cE(6,{\scriptstyle {3 \over 2}},{\scriptstyle {3 \over 2}};\tau)+\beta_6\,\cE(6,{\scriptstyle {5 \over 2}},{\scriptstyle {5 \over 2}};\tau)+\gamma_6\,\cE(6,{\scriptstyle {7 \over 2}},{\scriptstyle {3 \over 2}};\tau)]\Big]\,.
}
From \cite{FKRinprogress}, we find that the contributions from the Ramanujan cusp form to the relevant generalised Eisenstein series  are 
\begin{align}
\cE(6,\threeh,\threeh;\tau)\big\vert_{\rm cusp} &\nn=   \frac{2 \Lambda(\Delta_{12};6)^2}{\pi \langle \Delta_{12},\Delta_{12}\rangle}\times H_{\Delta_{12}}(\tau)\,,\\
\cE(6,\fiveh,\fiveh;\tau)\big\vert_{\rm cusp} &\label{eq:cuspEx}= -  \frac{128 \Lambda(\Delta_{12};6)^2}{225\pi \langle \Delta_{12},\Delta_{12}\rangle}\times H_{\Delta_{12}}(\tau)\,,\\
\cE(6,\threeh,\sevenh;\tau)\big\vert_{\rm cusp} &\nn=   \frac{32 \Lambda(\Delta_{12};6)^2}{105\pi \langle \Delta_{12},\Delta_{12}\rangle}\times H_{\Delta_{12}}(\tau)\,,
\end{align}
where we have used a classic result due to Manin \cite{Manin:1973}, which implies that 
\begin{equation}
\Lambda(\Delta_{12};6):\Lambda(\Delta_{12};8):\Lambda(\Delta_{12};10) = 1:\frac{5}{4}:\frac{12}{5}\,,
\end{equation} 
so that all  the numerators are expressed in terms of $\Lambda(\Delta_{12};6)$, 
The coefficients in \eqref{N3} were determined in \cite{Chester:2020vyz}, to be
\begin{equation}
\quad\alpha_6 =\frac{135}{52 }\, ,  \qquad \beta_6 =\frac{30375}{832  } \, , \qquad \gamma_6=\frac{42525}{832  }  \,.
\end{equation}
With these particular coefficients we find that the sum of the cuspidal contributions \eqref{eq:cuspEx} to \eqref{N3} is zero.

The cancellation of the cusp forms contributions is indeed expected, since they do not appear in the localisation computation. So  for each $s$, they must cancel between the different $\cE(s,s_1,s_2;\tau)$ that appear in $\cH_N^i(\tau)$, as we demonstrated  explicitly  for the  above  example. This property leads to the aforementioned constraints on the coefficients $\alpha_{j,\ell,r,s}$. We have verified and exploited similar cancellations for all the other generalised Eisenstein series that appear in the large-$N$ expansion of $\cH_N^i(\tau)$  up to  order $O(N^{-7})$ and explicit results can be found in the attached \texttt{Mathematica} file.

\subsection{Lattice sum representation}
\label{sec:lattice}

\begin{figure}[h]
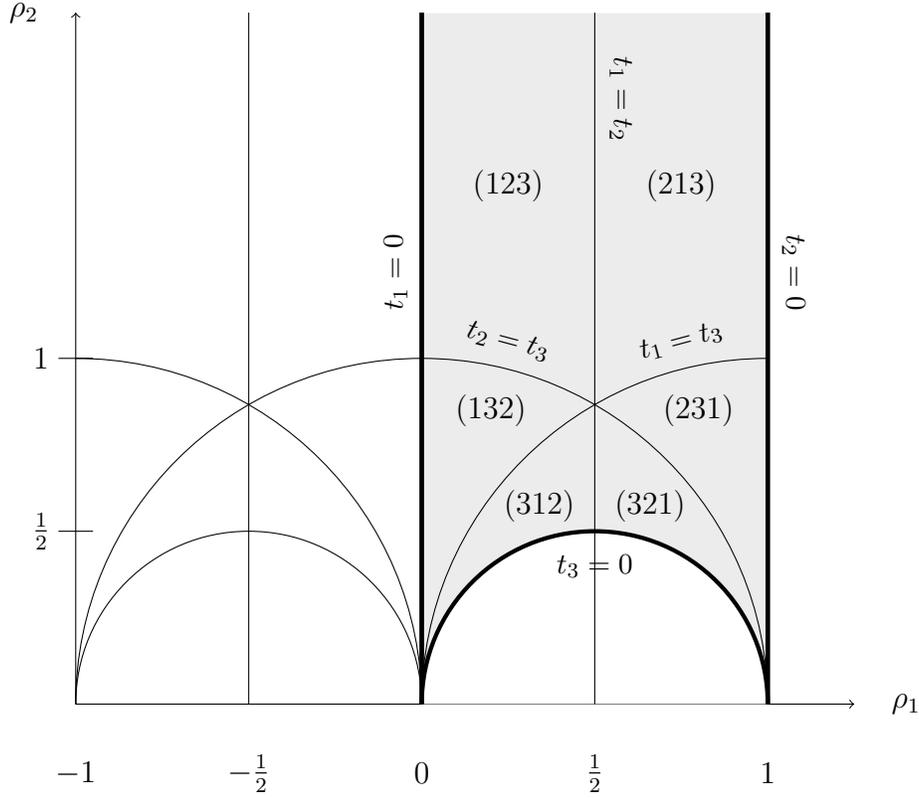

\begin{center}
\tikzpicture[scale=2.3]
\scope[xshift=-3cm,yshift=0.0]
\draw[white,fill=gray!15] (0,0) rectangle (2,4.0);
\draw (-2,0) -- (2,0) ;
\draw  [ultra thick] (0,0) -- (0,4.) ;
\draw   [ultra thick](2,0) -- (2,4) ;
\draw [->] (2,0) -- (2.5,0);
\draw (2.8, 0.0) node{$\ttau_1$};
\draw [->] (-2,0) -- (-2,4.0) ;
\draw  (-2.3, 4.0) node{$\ttau_2$};
\draw [fill=white,ultra thick] (2,0) arc(0:180:1.0) ;
\draw  (2,0) arc(0:90:2.0) ;
\draw (0,0) arc(180:90:2.0) ;
\draw (1,0) -- (1,4) ;
\draw (-2.1,1) -- (-1.9,1) ;
\draw (-2.1,2) -- (-1.9,2) ;
\draw  [ultra thick] (0,0) -- (0,4.) ;
\draw   [ultra thick](2,0) -- (2,4) ;
\draw (0,0) arc(0:180:1.0) ;
\draw  (0,0) arc(00:90:2.0) ;
\draw (-2,0) arc(180:90:2.0) ;
\draw (-1,0) -- (-1,4) ;
%
\draw (0, -0.4) node{$0$};
\draw (1, -0.4) node{$\frac{1}{2}$};
\draw (2, -0.4) node{$1$};
\draw (-1, -0.4) node{$-\frac{1}{2}$};
\draw (-2, -0.4) node{$-1$};

\draw (-2.2, 1.0) node{$\frac{1}{2}$};
\draw (-2.2, 2.0) node{$1$};

\draw (0.5, 3) node{{$(123)$}};
\draw (1.5, 3) node{{$(213)$}};
\draw (0.4, 1.7) node{{$(132)$}};
\draw (1.6, 1.7) node{{$(231)$}};
\draw (0.68, 1.15) node{{$(312)$}};
\draw (1.32, 1.15) node{{$(321)$}};
\draw (-0.15, 2.5) node[rotate=90]{\small $t_1=0$};
\draw (1.15, 3.5) node[rotate=-90]{\small $t_1=t_2$};
\draw (2.15, 2.5) node[rotate=-90]{\small $t_2=0$};
\draw (0.5, 2.1) node[rotate=-15]{\small $t_2=t_3$};
\draw (1.5, 2.1) node[rotate=15]{\small $t_1=t_3$};
\draw (1, 0.8) node{\small $t_3=0$};
\endscope

\endtikzpicture
\end{center} 
\caption{In the $\rho$-plane, the domain $t_i>0$ is mapped to the region shaded in grey, isomorphic to a fundamental domain of $\Gamma_0(2)$.  Here the labels $(ijk)$ denote the ordering $t_i<t_j<t_k$ of the Schwinger parameters.}
\label{fig:fundomain}
\end{figure}

The linear relations between the coefficients $\alpha_{j,\ell,r,s}$ in \eqref{eq:Hni} that lead to the cancellation of cusp forms suggests that there is a more refined basis of modular functions than the $\cE(s,s_1,s_2;\tau)$. Following \cite{Green:2008bf}, which we summarise in appendix \ref{sec:geneis}, 
we introduce a new class of modular invariant objects defined via lattice sums over four integers
\begin{align}
\cE^w_{i,j}(\tau\btau) \equiv &\sum_{\underset {p_1+p_2+p_3=0} {p_1,p_2, p_3\ne 0}} \int_0^\infty d^3 t\,B_{i,j}^w(t)\,  \exp\Big( - \frac{\pi}{\tau_2}\sum_{i=1}^3 t_i |p_i|^2  \Big)\,,
\label{eq:lattint}
\end{align}
 where $p_i=m_i+n_i\tau$ with $m_i,n_i\in \ZZ$.
The function $B_{i,j}^w(t)$ is a symmetric function of $(t_1,t_2,t_3)$ that can be expressed in  the form
\begin{equation}
B_{i,j}^w(t) \equiv Y(t)^{\frac{w-3}{2}} A_{i,j}(\ttau(t))\,,
\label{eq:Bijw}
\end{equation}
where $Y(t) = t_1 t_2+t_1t_3 +t_2t_3 $ and
\begin{align}
 \ttau_1(t) = \frac{t_1}{t_1+t_2} \,,\qquad \ttau_2(t) =\frac{\sqrt{Y(t)}}{t_1+t_2}\,, \qquad {\rm with} \qquad  \ttau(t) = \ttau_1(t)+i \ttau_2(t) \, .
\label{eq:rhodef}
\end{align}
 The domain $t_i \ge 0$ translates into the domain $0\le Y(t)\le  \infty$ and
\bea
 0\le \rho_1 \le 1\,, \ \ \qquad (\rho_1-\half)^2 + \rho_2^2 \ge \frac{1}{4}\,, 
\label{eq:domain} 
 \eea
 which is the fundamental domain of $\Gamma_0(2)$ illustrated in figure \ref{fig:fundomain}.

The function  $A_{1,0}(\ttau)$ originally arose in the construction of the $d^6R^4$ coefficient $\cE(4, \frac{3}{2},\frac{3}{2};\stau)$  in  \cite{Green:2005ba}, and a more general discussion in  \cite{Green:2008bf} presented the general procedure for constructing linear combinations of $\cE(s,s_1,s_2;\stau)$. 
 The expressions for $A_{i,j}(\ttau)$ are Laurent polynomials in $\ttau_2$ with coefficients which are polynomial in $\ttau_1$.  They  satisfy  homogeneous Laplace equation with respect to the Laplace-Beltrami operator $\Delta_\ttau = \ttau_2^2(\partial_{\ttau_1}^2 +\partial_{\ttau_2}^2)$
\begin{equation}
\label{DeltaOm}
\left[\Delta_\ttau   - s(s-1) \right] \, A_{i,j}  (\ttau)=0\, ,
\end{equation}
inside the domain\footnote{{The upper half $\rho$-plane is denoted by $\mathcal{H}=\{\ttau \in \mathbb{C}\,\vert\,\ttau_2>0\}$, while the congruence subgroup $\Gamma_0(2)$ is defined as $\Gamma_0(2)=\{\left(\begin{smallmatrix} a& b \\ c & d\end{smallmatrix}\right)\in SL(2,\mathbb{Z})\,\vert\, c \equiv 0\,\rm{mod}\,2\}$.}} $\ttau \in \mathcal{H}\backslash \Gamma_0(2)$ and where $s=3i+j+1$ (for example, $s=4$ for  $A_{1,0}$).

 In appendix \ref{eq:aijdef} we present the systematic construction of $A_{i,j}$ given by Don Zagier  in unpublished notes that are expanded on in section 5.2 of  \cite{DHoker:2018mys}, where they are called `modular local polynomials', see also \cite{Bringmann}.

When translated into $t$-variables \eqref{DeltaOm} implies that the function $B_{i,j}^w(t)$ defined in \eqref{eq:Bijw} satisfies 
the homogeneous Laplace equation
 \begin{equation}
\label{DeltaT}
\left[\Delta_t   - s(s-1) \right] \, B^w_{i,j}  (t)=0\,,
\end{equation}
where again $s=3i+j+1$ and the laplacian $\Delta_t$ {is defined by}
\begin{align}\label{eq:deltat}
\Delta_t [F(t) ] = -2\sum_{i=1}^3 \partial_i[ t_i \, F(t)\, ] +\sum_{i,j=1}^3  \partial_i \partial_j  \{  [t_i t_j + (2\delta_{ij} -1) Y(t)] \,F(t)\}\, .
\end{align}

Furthermore it is straightforward to show that
\begin{equation}\label{eq:DeltaTauOm}
 \Delta_\stau \exp\Big( - \frac{\pi}{\tau_2}\sum_{i=1}^3 t_i |p_i|^2  \Big) = \Delta_t \exp\Big( - \frac{\pi}{\tau_2}\sum_{i=1}^3 t_i |p_i|^2  \Big)\,,
\end{equation}
so by  acting with $\Delta_\stau$ on \eqref{eq:lattint} and combining \eqref{eq:DeltaTauOm} and  \eqref{DeltaT} we deduce
\begin{equation}
\big[ \Delta_\stau - s(s-1) \big] \cE^w_{i,j}(\stau) = {\mbox{boundary terms}}\,,
\end{equation}
where the boundary terms arise from integrating by parts the Laplacian $\Delta_t$. In appendix \ref{app:LapEig}, we show that the boundary terms produce the ``source terms'' in the inhomogeneous Laplace eigenvalue equation, 
\begin{align}
[\Delta_\stau -s(s-1)] \cE_{i,j}^{w}(\stau) = \sum_{\delta=\frac{2-s}{2}}^{\frac{s-2}{2}} c_{i,j}(w,\delta) E\left(\frac{w}{2}+\delta;\stau \right) E \left(\frac{w}{2}-\delta;\stau \right) +d_{i,j}(w) E(w;\stau)\,,\label{eq:LapcEijw}
\end{align}
for particular values of the coefficients $c_{i,j}(w,\delta), \, d_{i,j}(w)$. 

We note a few important facts. 
Firstly, when the weight $w$ and the eigenvalue $s=3i+j+1$ have opposite parity (as in our case), the source terms are  bilinears in Eisenstein series with half-integer indices.
Secondly, the modular invariant function $\cE_{i,j}^{w}(\stau)$ satisfies an inhomogeneous Laplace equation where the source terms have { total ``trascendental weight'' given by $w =s_1+s_2$. }
It is always possible to use the functional equation 
\begin{equation}
\Gamma(s) E(s;\tau) = \Gamma(1-s)E(1-s;\tau)\,,
\end{equation}
to rewrite \eqref{eq:LapcEijw} so that all Eisenstein series in the source term appear with positive indices, with the drawback of spoiling uniform trascendentality in weight.

 Lastly, since we are interested in eigenvalues $s\geq w+1$ it is always possible to use \eqref{LaplaceEq} and invert the laplacian\footnote{{We can check from the lattice sum representation \eqref{eq:lattint} that for the range of eigenvalues of present interest, the coefficient of the homogeneous solution $E(s;\tau)$ vanishes. This implies that there is no issue in inverting the Laplace operator in \eqref{eq:LapcEijw}.}} over the single Eisenstein series that appear as source terms to get 
\begin{equation}
\cE_{i,j}^{w}(\stau) = \sum_{\delta=\frac{2-s}{2}}^{\frac{s-2}{2}} c_{i,j}(w,\delta) \cE\! \left(s,\frac{w}{2}+\delta,\frac{w}{2}-\delta;\stau \right) +\frac{d_{i,j}(w)}{w(w-1)-s(s-1)} E(w;\stau)\,.
\end{equation}

Note that for eigenvalue $s$ such that ${\rm dim}\,\cS_{2s}=0$ it is possible to find an isomorphism between the two vector spaces spanned by either $\cE_{i,j}^w(\stau)$ or by $\cE(s,s_1,s_2;\stau)$ with $s=3i+j+1$ and fixed weight $w=s_1+s_2$ (modulo the addition of single Eisenstein series). 
For example, in the first few cases relevant to \eqref{eq:HN3} up to order $O(N^{-2})$  we have 
\begin{align}
\cE_{1,0}^{3}(\stau) &\nn= -\frac{9}{10}\pi \cE(4,\threeh,\threeh;\tau) -\frac{3}{10}\pi E(3;\tau)\,,\\
\cE_{1,1}^{4}(\stau) &= -\frac{81}{28}\pi \cE(5,\threeh,\fiveh;\tau) -\frac{99}{140}\pi E(4;\tau)\,,\\
\cE_{2,0}^{4}(\stau) &\nn= -\frac{15}{11}\pi \cE(7,\threeh,\fiveh;\tau) -\frac{1}{14}\pi E(4;\tau)\,.
\end{align}
We can use these relations and similar ones at higher order in $1/N$ to express $ \cH^i_N(\tau)$ in terms of $ \cE^{w}_{i,j}(\tau)$ (and $E(s;\tau)$). For example, up to $1/N^3$, we have  
\begin{align}
 \cH^i_N(\tau)\big\vert_{N^{-1}} &\label{eq:Eijw1}=-\frac{15}{4\pi} \cE^{3}_{1,0}(\tau) -\frac{9}{8} E(3;\tau)\,,\\
 \cH^i_N(\tau)\big\vert_{N^{-2}} &\label{eq:Eijw2} = C_1+\frac{945}{64\pi} \cE_{2,0}^{4}(\tau) -\frac{105}{22\pi} \cE^{4}_{1,1}(\tau) - \frac{297}{128} E(4;\tau)\,,\\
   \cH^i_N(\tau)\big  \vert_{N^{-3}}  &\label{eq:Eijw3} = \frac{645}{572 \pi} \cE^{3}_{1,0}(\tau) -\Big[\frac{945}{104 \pi}\cE^5_{1,2}(\tau)+\frac{945}{208 \pi} \cE^3_{1,2}(\tau)- \frac{63}{104\pi} \cE^3_{0,5}(\tau)\Big]\\
  &\nn\!\!\!\!\!\!\!\!+\Big[\frac{31185}{544\pi}\cE^5_{2,1}(\tau) +\frac{23625}{544 \pi} \cE^{3}_{2,1}(\tau) -\frac{23625}{4352\pi} \cE^{3}_{1,4}(\tau)\Big]\\
  &\nn\!\!\!\!\!\!\!\!-\Big[\frac{45045}{512\pi} \cE^{5}_{3,0}(\tau) + \frac{183645}{2048 \pi}\cE^{3}_{3,0}(\tau) -\frac{10395}{1024\pi} \cE^{3}_{2,3}(\tau)\Big] -\frac{496125}{146432} E(3;\tau) + \frac{5805}{512} E(5;\tau)\,.
\end{align}
We thus see that $\cH_N^i(\tau)$  can be written in terms of the more refined basis of modular functions $\cE^w_{i,j}(\tau)$ and $E(w;\tau)$, which manifestly does not include the unwanted cusp form contributions that appear in generic sums of  $\cE(s,s_1,s_2;\tau)$. Furthermore, they are the objects that are naturally written as lattice sums. This pattern generalises to all the higher orders, which we have verified explicitly up to order $O(N^{-7})$.  
The explicit expression for the lattice sum representation is described in some detail in appendix \ref{sec:geneis}.

\section{Numerical estimate of free energy at finite $N$}
\label{sec:finiteN}

We will now discuss how the large-$N$ finite-$\tau$ results for $\cC_N$ and $\cH_N$ can be used to accurately estimate these quantities at finite $N$ and $\tau$. This is especially important for $\cH_N$, for which we have no exact expression.  These finite-$N$ numerical results will provide an important input for the numerical bootstrap study of $\mathcal{N}=4$ SYM following \cite{Chester:2021aun}. 

Our strategy is to divide $\cC_N$ and $\cH_N$ into perturbative and non-perturbative terms as $g_{_{YM}}^2
\to 0$, or equivalently for $\tau_2\gg1$, arising in the matrix model \eqref{ZFull} and \eqref{ZInstSum} 
\es{divideHG}{
\cC_N=\cC_N\big\vert_{pert}+\cC_N\big\vert_{non-pert}\,,\qquad\qquad\qquad \cH_N=\cH_N\big\vert_{pert}+\cH_N\big\vert_{non-pert}\,, 
}
which is a meaningful distinction for all $N$. The perturbative sector only depend on $\tau_2$, and was already computed for finite $N$ and $\tau_2$ in \cite{Chester:2019pvm,Chester:2020dja} using the method of orthogonal polynomials. For instance, for $\cC_N$ we have \cite{Chester:2019pvm}\footnote{A slightly simplified version of this formula can be found in (A.12) in  \cite{Dorigoni:2022cua}.}
\es{GHdiv}{
\cC_N\big\vert_{pert}=-\tau_2^2\partial_{\tau_2}^2\int_0^\infty dw\frac{e^{-\frac{w^2}{\pi\tau_2}}}{2\sinh^2 w}\Big[L_{N-1}^{(1)}({\scriptstyle\frac{w^2}{\pi\tau_2}})-\sum_{i,j=1}^N(-1)^{i-j}L_{i-1}^{(j-i)}({\scriptstyle\frac{w^2}{\pi\tau_2}})L_{j-1}^{(i-j)}({\scriptstyle\frac{w^2}{\pi\tau_2}})\Big]\,,
}
where $L_i^{(j)}(x)$ are generalised Laguerre polynomials, and note that for all $N$ we need only perform a single integral and some finite sums. The analogous expression for $\cH_N\vert_{pert}$ from \cite{Chester:2020dja} involves two integrals and some finite sums for all $N$. Since it is much more complicated, we do not show it here explicitly, but we include it in the attached \texttt{Mathematica} file.

For the non-perturbative sector, we will use the large-$N$ finite-$\tau$ expressions from the previous sections, which were written in terms of Eisenstein series and generalised Eisenstein series. The advantage of only using the non-perturbative terms of these functions is that we avoid the divergence at the free theory point $\tau_2\to \infty$ that occurs for their perturbative terms, as seen from the Fourier expansions \eqref{EisensteinExpansion} and \eqref{genEisPert}. From the Fourier decomposition of the non-holomorphic Eisenstein series \eqref{EisensteinExpansion} we see that non-perturbative terms correspond to Fourier modes with $k\neq 0$ while from \eqref{genEsketch} we see that the non-perturbative terms of the generalised Eisenstein series correspond to terms with $(n,m)\neq (0,0)$. The non-perturbative sectors converge extremely quickly for $\tau$ in the fundamental domain
\es{fund}{
|\tau|\geq1\,,\qquad |\Re(\tau)|\leq\frac12\,,
}
so that only a few values of $k\neq 0$ and $(n,m)\neq(0,0)$ need be included from the Fourier expansions to achieve good accuracy. Since the large-$N$ expansion is an asymptotic expansion, we expect that the expansion will actually get worse at some order. We observe that the best approximation is given by keeping terms up to $O(1/N^{\frac{5}{2}})$ in the non-perturbative sector 
for $\cC_N$,\footnote{For $\cC_N(\tau\btau)$ we may use the exact result in  \cite{Dorigoni:2021guq}. However, as we will show, the simple approximation given here already provides an extremely good estimate. And in this way, we treat $\cC_N(\tau\btau)$ and $\cH_N(\tau\btau)$ uniformly.} 
\es{GApprox}{
\cC_N\big\vert_{non-pert}&\approx \Bigg[
 -\frac{3\sqrt{N}}{2^4 } E( {\scriptstyle {3 \over 2}};\tau\btau)+\frac{45}{2^8 \sqrt{N}}E( {\scriptstyle {5 \over 2}};\tau\btau)+\frac{1}{{N}^{\frac32}}\big[-\frac{39}{2^{13} }E( {\scriptstyle {3 \over 2}};\tau\btau)+\frac{4725}{2^{15}}E( {\scriptstyle {7 \over 2}};\tau\btau)\big]\\
&\qquad+\frac{1}{{N}^{\frac52}}\big[-\frac{1125}{2^{16} }E( {\scriptstyle {5 \over 2}};\tau\btau)+\frac{99225}{2^{18} }E( {\scriptstyle {9 \over 2}};\tau\btau)\big]\Bigg]_{k\neq 0}\,,\\
}
and for $\cH_N$: 
\es{HApprox}{
& \cH_N\big\vert_{non-pert} \approx \Bigg[6\sqrt{N} E( {\scriptstyle {3 \over 2}};\tau\btau)-\frac{9}{2\sqrt{N}}E( {\scriptstyle {5 \over 2}};\tau\btau)
+\frac{27}{2^3 N}\cE(4,{\scriptstyle {3 \over 2}},{\scriptstyle {3 \over 2}};\tau\btau)\\
+& \, \frac{1}{{N}^{\frac32}}\big[\frac{117}{2^8 }E( {\scriptstyle {3 \over 2}};\tau\btau)-\frac{3375}{2^{10} }E( {\scriptstyle {7 \over 2}};\tau\btau)\big]+\frac{1}{N^2}\big[- \frac{14175}{704}\cE(7,{\scriptstyle {5 \over 2}},{\scriptstyle {3 \over 2}};\tau\btau) + \frac{1215}{88}\cE(5,{\scriptstyle {5 \over 2}},{\scriptstyle {3 \over 2}};\tau\btau)\big]\\
+& \, \frac{1}{{N}^{\frac52}}\big[\frac{675}{2^{10} }E( {\scriptstyle {5 \over 2}};\tau\btau)-\frac{33075}{2^{12} }E( {\scriptstyle {9 \over 2}};\tau\btau)\big]
\Bigg]_{\substack{ \!\!\!\!\!\!\!\!\!\!\!\!\!\!\!\!\!\! k\neq 0\\ (n,m)\neq(0,0)}}\,,
}
where on the RHS we only keep the $k\neq 0$ instanton terms in the Fourier expansions for $E(s;\tau\btau)$ in \eqref{EisensteinExpansion} and the instanton/anti-instanton sectors $(n,m)\neq(0,0)$ for $\cE(s,s_1,s_2;\tau\btau)$ given in\eqref{genEsketch}. The explicit expressions are given in the attached \texttt{Mathematica} file.

 \begin{figure}[]
\begin{center}
       \includegraphics[width=.49\textwidth]{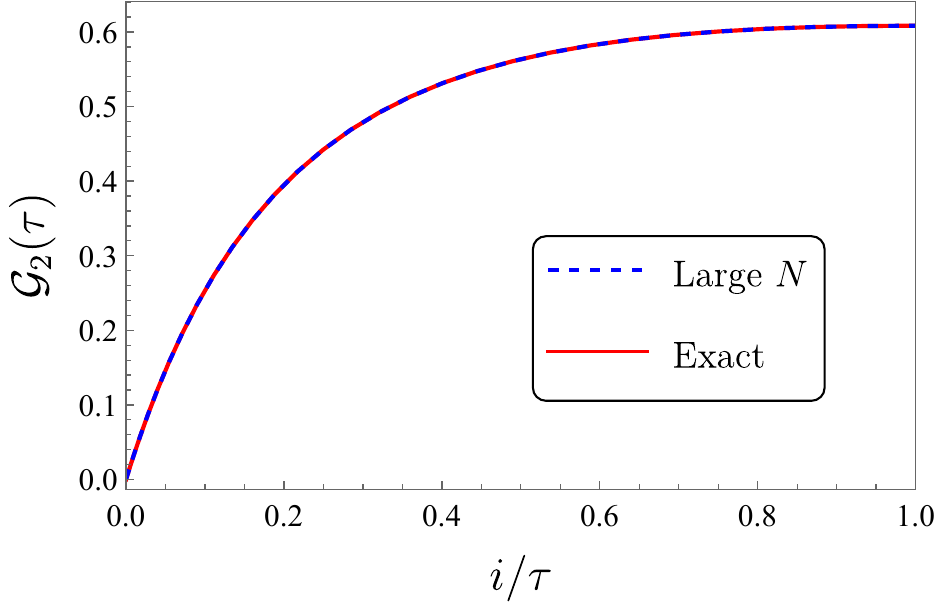}
      \includegraphics[width=.49\textwidth]{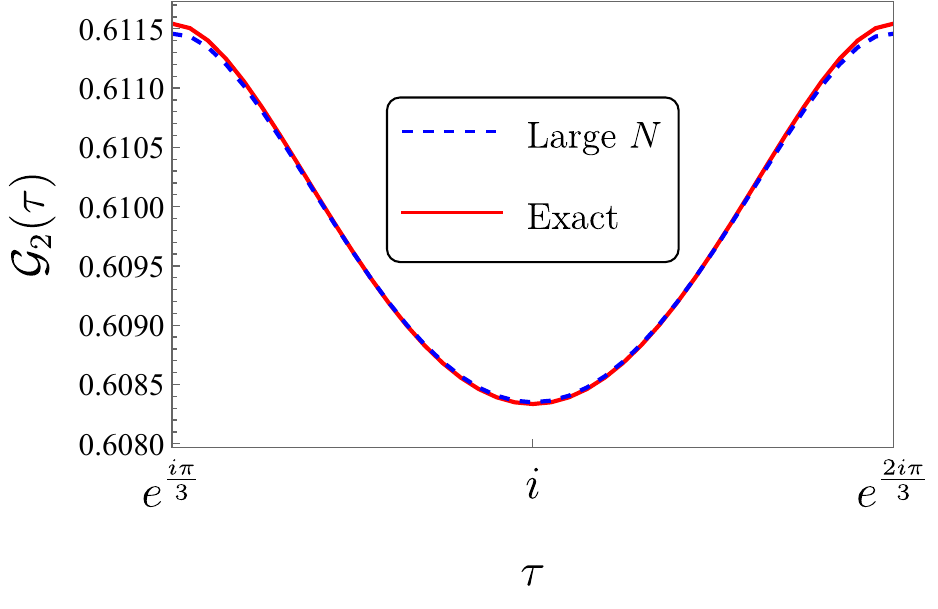}
         \includegraphics[width=.49\textwidth]{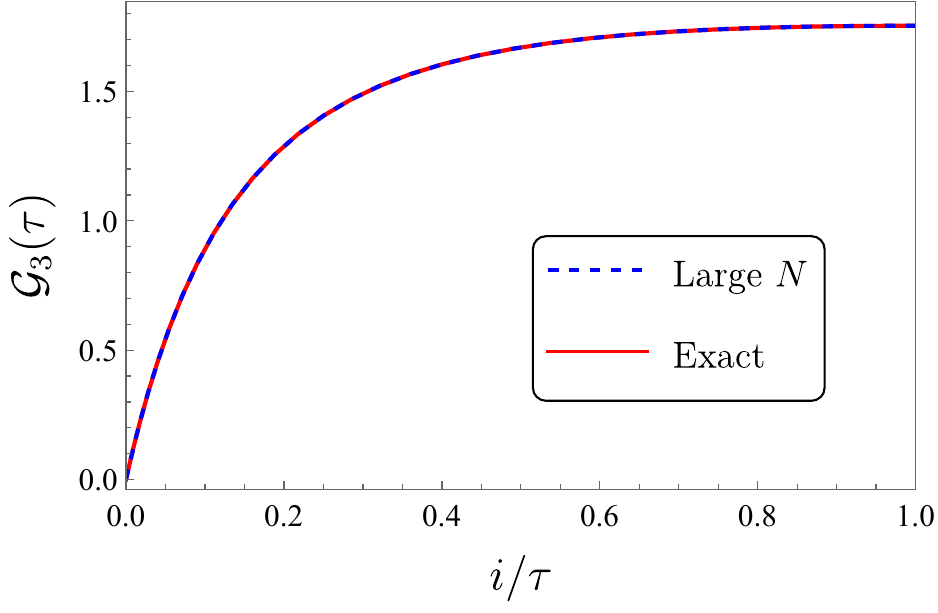}
      \includegraphics[width=.49\textwidth]{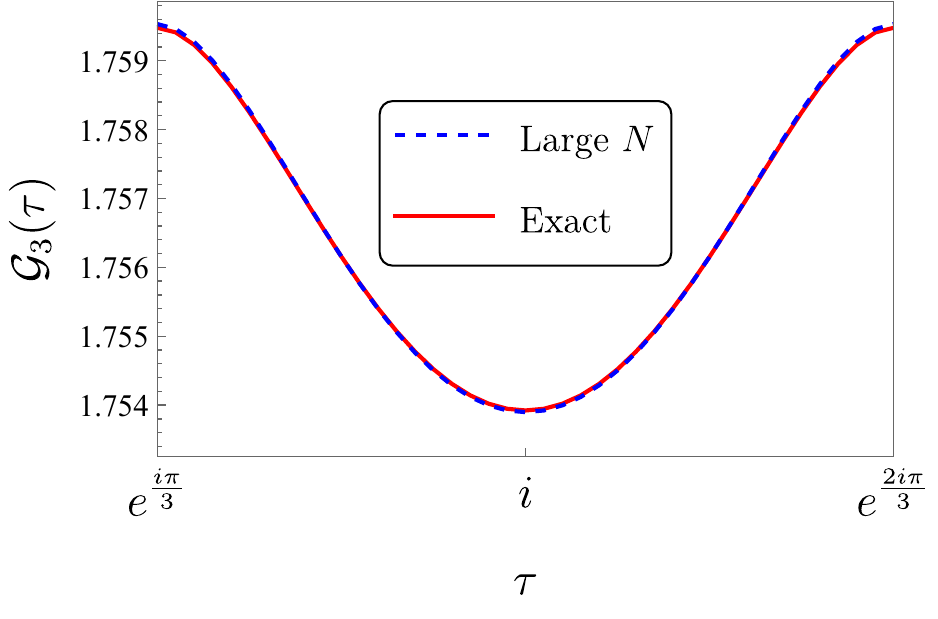}
             \includegraphics[width=.49\textwidth]{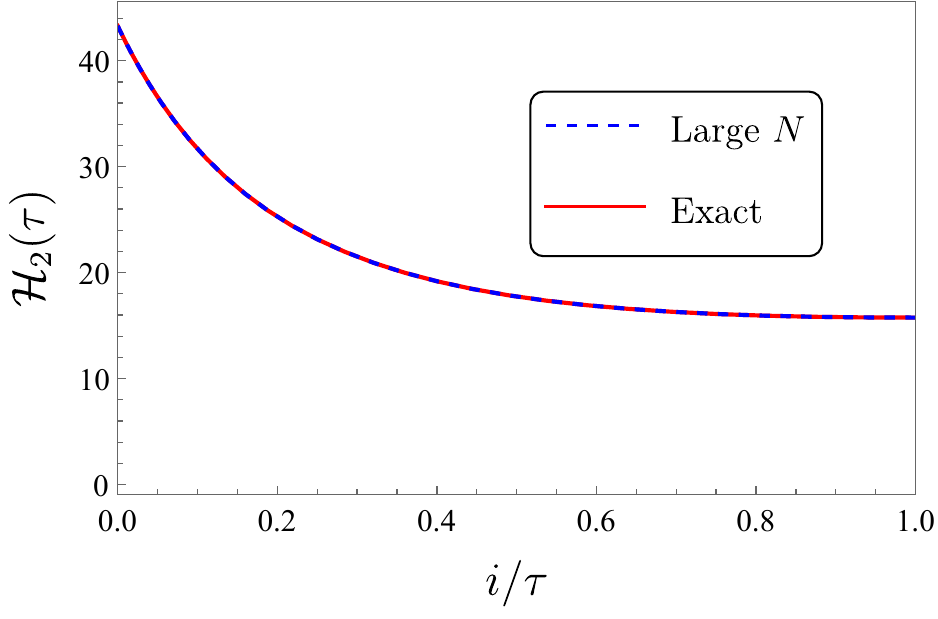}
      \includegraphics[width=.49\textwidth]{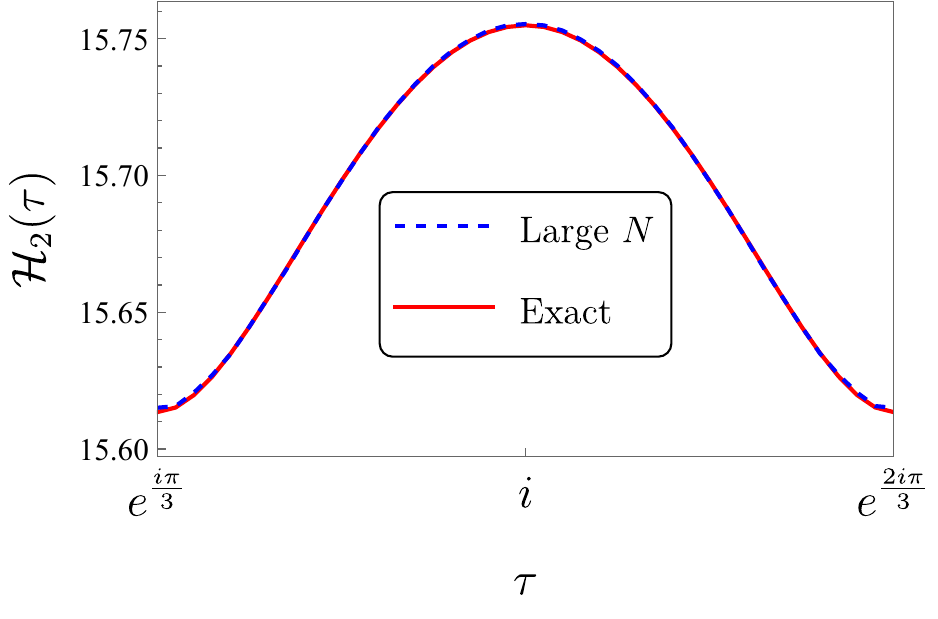}
         \includegraphics[width=.49\textwidth]{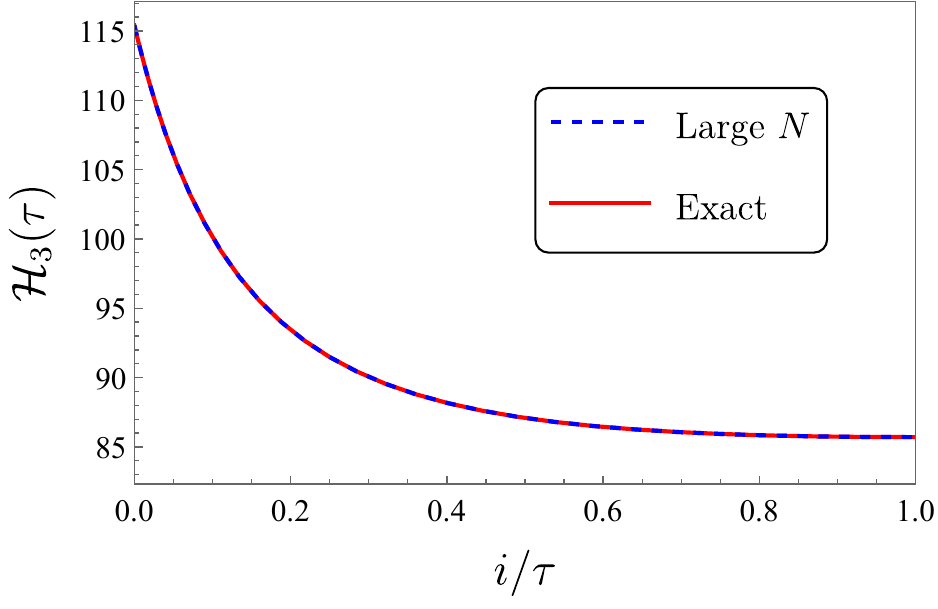}
      \includegraphics[width=.49\textwidth]{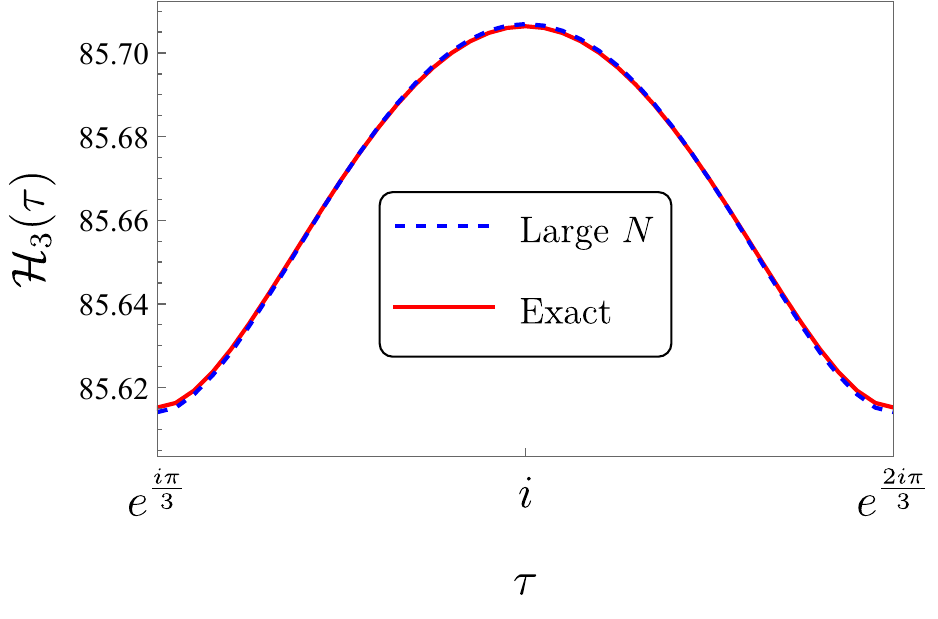}
	\caption{Comparison of large-$N$ approximation to exact $N$ for $\cC_N$ and $\cH_N$ for two paths in the fundamental domain of $\tau$: from the free theory $\tau\to i\infty$ to the $\mathbb{Z}_2$ self-dual point $\tau=i$ along the line $\tau_1=0$, and from $\tau=i$ to the $\mathbb{Z}_3$ self-dual point $\tau=e^{i\pi/3}$ along the arc $|\tau|=1$.}
\label{compare}
\end{center}
\end{figure}  
 
We can check our approximations for $\cC_N$ and $\cH_N$ by comparing to low values of $N$, where the exact expression can be computed from the matrix model integral \eqref{ZFull} by doing $N-1$ integrals and truncating the instanton expansion in \eqref{ZInstSum}, which converges very quickly in the fundamental domain. This finite-$N$ calculation was performed for $N=2,3$ in \cite{Chester:2021aun}. In Figure \ref{compare}, we compare these finite-$N$ expressions to our large-$N$ approximation for two paths in a fundamental domain of $\tau$: from the free theory $\tau\to i\infty$ to the $\mathbb{Z}_2$ self-dual point $\tau=i$ with $\tau_1=0$, and from $\tau=i$ to the $\mathbb{Z}_3$ self-dual point $\tau=e^{i\pi/3}$ along the arc $|\tau|=1$. Along the first path, instantons are very small, so is perhaps not so surprising that our large-$N$ approximation is so accurate, as we have included the exact perturbative sector. The precise match is more surprising along the second path, where the range of the plots is much smaller, and the variation is mostly due to instantons. The greatest discrepancy between large and finite $N$ is at $\tau=e^{i\pi/3}$, but even that discrepancy is only around $.01\%$ relative error, and this error decreases as we increase from $N=2$ to $N=3$. We are thus confident that our approximations are very accurate for all $N$ and $\tau$.

\section{Constraints on stress tensor correlator}
\label{sec:correlator} 

We now show how the stress tensor correlator can be constrained at large $N$ using the localisation quantities computed in the previous sections. We consider the four point function of the bottom component $S$ of the stress tensor multiplet, often also denoted by $\mathcal{O}_2$.  This operator is a dimension 2 scalar in the ${\bf 20}'$ of $SO(6)_R$, and can thus be represented as a rank-two traceless symmetric tensor $S_{IJ}(\vec{x})$, with indices $I, J = 1, \ldots, 6$.
For simplicity we will contract these indices with  polarisation vectors $Y^I$, with $Y \cdot Y = 0$.  The four-point function  is then fixed by superconformal symmetry to take the form \cite{Eden:2000bk, Nirschl:2004pa}
 \es{FourPoint}{
  \langle S(\vec{x}_1, Y_1) S(\vec{x}_2, Y_2) S(\vec{x}_3, Y_3)  S(\vec{x}_4, Y_4) \rangle 
   = \frac{1}{\vec{x}_{12}^4 \vec{x}_{34}^4}
    \left[ 
     \vec{\cS}_\text{free} + {\cal T}(U, V) \vec{\Theta} 
    \right] \cdot \vec{\cB} \,,
 }
where $\vec{x}_{ij}  \equiv \vec{x}_i - \vec{x}_j$, and $U \equiv \frac{ \vec{x}_{12}^2 \vec{x}_{34}^2}{ \vec{x}_{13}^2 \vec{x}_{24}^2}$ and $V\equiv \frac{ \vec{x}_{14}^2 \vec{x}_{23}^2}{ \vec{x}_{13}^2 \vec{x}_{24}^2}$ are the usual conformal invariant cross-ratios. The precise forms of $\vec{\cS}_\text{free}$, $\vec{\Theta}$ and $\vec{\cal B}$  can be found in \cite{Binder:2019jwn}, and all non-trivial information is given by the $R$-symmetry invariant correlator $\mathcal{T}(U,V)$, which will be our focus. 

We can  expand $\mathcal{T}(U,V)$ at large $c = (N^2 - 1)/4$, whose functional form is fixed by the analytic bootstrap \cite{Rastelli:2017udc,Alday:2014tsa, Chester:2020vyz} to take the form 
\es{TlargeN}{
\cT=\frac{8}{c}\cT^{R}+\frac{1}{c^2}\big[ \cT^{R|R}+B_0\cT^0 \big]+\frac{1}{c^3}\big[ \cT^{R|R|R}+\sum_{i=0,2,3,4}\overline{B_i}\cT^i \big]+\dots\, ,
}
where the ellipses denote both higher loop terms as well as contact terms involving higher derivative terms such as $R^4$, which at large $c$ and finite $\tau$ generically come with fractional powers of $c$.\footnote{ The first few higher derivative corrections to the correlator, including $R^4, d^4R^4$ and $d^6R^4$, were determined in \cite{Chester:2019jas,Chester:2020vyz}. Furthermore, starting from order $O(1/c^3)$ there is actually no difference between the loop and higher derivative expansion in the large-$c$ finite-$\tau$ limit, since both $d^{12}R^4$ and $\cT^{R|R|R}$ are the  same order in derivatives. In this case, we simply define the terms written in \eqref{TlargeN} to be those that are independent of $\tau$, as higher derivative terms generically depend on $\tau$.} 
We denote the supergravity exchange term by $\cT^R$, which is given by \cite{Arutyunov:2000py}
\es{SintM}{
 \mathcal{T}^R= -\frac18U^2\bar D_{2,4,2,2}(U,V)\,,
}
where $\bar D_{a,b,c,d}(U,V)$ are standard functions whose explicit form can be found e.g. in \cite{Dolan:2011dv}. The coefficient of $\cT^R$ is fixed by the conformal Ward identity \cite{Osborn:1993cr}. Next we have the one-loop graviton exchange term $ \cT^{R|R}$, which was fixed in position space in \cite{Aprile:2017bgs} from tree-level data using the AdS unitarity method of \cite{Aharony:2016dwx} up to a contact-term ambiguity 
\es{T0}{
\mathcal{T}^0=U^2\bar D_{4,4,4,4}\,.
}
In general, $\cT^n$ for integer $n$ denotes contact terms with $2n+8$ derivatives, which are related to the divergences that appear at each loop level. The two-loop graviton exchange term is denoted by $ \cT^{R|R|R}$, and was conjecturally fixed in \cite{Huang:2021xws,Drummond:2022dxw} up to four contact-term ambiguities, whose explicit form we will not use.  The coefficients of contact-term ambiguities (i.e.~$B_0$ and $\overline{B_i}$) are not fixed by general analytic bootstrap constraints, and so require the supersymmetric localisation constraints discussed here.

We can constrain $\cT(U, V)$ using an exact property of $\cH_N(\tau\btau)$ \cite{Chester:2020dja}: 
\es{int}{
 \cH_N(\tau\btau) &= 48 \zeta(3) c+c^2 I_4[\cT]\,,\\
     I_4[\cT]&\equiv\frac{32}{\pi}  \int_0^\infty \! dr\!\int_0^{2\pi} d\theta\, r^3 \sin^2 \theta \, \frac{1 + U + V}{U^2} \bar{D}_{1,1,1,1}(U,V) \cT(U, V) \bigg|_{\substack{U = 1 + r^2 - 2 r \cos \theta \\
    V = r^2}}\,,
}
where the $48\zeta(3)c$ term comes from the free theory contribution to \eqref{FourPoint}. A similar integrated constraint  between $\cT(U, V)$ and $\cC_N(\tau)$ was derived in \cite{Binder:2019jwn} and  has been studied in the literature so we will not use its detailed form here. Since we have two integrated constraints, one from $\cC_N$ and one from $\cH_N$, that means we can fix at most two $B$'s at each order in $1/c$ in \eqref{TlargeN}. There is just a single $B_0$ at $O(1/c^2)$, which was fixed in \cite{Chester:2019pvm} using the $\cC_N$ integrated constraint to be
\es{B}{
B_0=\frac{15}{4}\,.
}
For the integrated constraint arising from $\cH_N$, we make use of the following integrals:
\es{ints}{
I_4[\cT^R]&=3-6\zeta(3)\,,\quad I_4[\cT^0]=\frac{16}{5}\,,\qquad I_4[\cT^{R|R}]=-\frac{48 \zeta (3)}{5}-\frac{83}{10}\,,\\
}
where the first two integrals were computed numerically in \cite{Chester:2020dja} using the explicit expressions in \eqref{SintM} and \eqref{T0}, while the last integral we compute now using the explicit expression in \cite{Aprile:2017bgs}. We find that these integrals applied to the correlator \eqref{TlargeN} using the integrated constraint \eqref{int} exactly match $\cH_N$ in \eqref{eq:HN} to one-loop order with the one-loop ambiguity $B_0$ fixed as in \eqref{B} and $C_0$ fixed in \eqref{Cs}. We do not have sufficient constraints yet to fix the ambiguities at higher loop orders, such as the four ambiguities at two loops \cite{Drummond:2022dxw, Huang:2021xws}, but the localisation data for $\cH_N$ is now available for that purpose.

\section{Discussion}
\label{sec:disc}

There are four main results of this paper. Firstly, we found recursion formulae that relate the half-integer terms and part of the integer terms in the large-$N$ expansion of $\cH_N(\tau)$ to $\cC_N(\tau)$. Secondly, we gave striking evidence that the large-$N$ expansion of $\cH_N(\tau)$ can be written as a lattice sum to any order in $1/N$. Thirdly, we used the large-$N$ expansion of $\cH_N(\tau)$ to verify the one-loop contact term in the stress tensor correlator as originally fixed using $\cC_N(\tau)$ in \cite{Chester:2019pvm}, and gave new constraints at higher loops. Lastly, we showed how the large-$N$ expansions of $\cH_N(\tau)$ and $\cC_N(\tau)$ can be used to accurately estimate these quantities for all finite $N$ and $\tau$. This estimate will be useful for the numerical bootstrap at finite $N$ and $\tau$ \cite{Chester:2021aun}, which was previously limited to low $N$ due to the difficulty of performing the $(N-1)$-dimensional matrix model integral.

While the recursion relations we found were sufficient to fix the half-integer powers of $N$ in $\cH_N(\tau)$ to all orders, we have not found a recursion relation that applies to the full $\cH_N(\tau)$, which could be used to compute $\cH_N(\tau)$ at finite $N$ and $\tau$ as was done for $\cC_N(\tau)$ in \cite{Dorigoni:2021guq}.  
Since $\cH_N(\tau)$ can be written as a lattice sum to all orders in $1/N$, it seems likely that a finite $N$ lattice sum expression should exist, analogous to the lattice sum expression for $\cC_N(\tau)$ in \eqref{eq:gsun}.  Importantly that expression is finite even though it contains the term in the lattice sum with $m=n=0$. Based on the lattice sum for the large-$N$ expansion of $\cH_N(\tau)$ in \eqref{eq:lattint}, a natural conjecture for the finite-$N$ expression would be 
\ie
\cH_N (\tau\btau) =\sum_{  \underset {p_1+p_2+p_3=0} {p_1,p_2, p_3\in\Z+\tau \Z}}  \int_0^\infty d^3t\, \cB_N (t_1,t_2,t_3)\, \exp\Big( - \frac{\pi}{\tau_2}\sum_{i=1}^3 t_i |p_i|^2  \Big) \,,
\label{eq:fullconj}
\fe
where the function $ \cB_N (t_1,t_2,t_3)$ is a symmetric function that can be shown to satisfy an inversion relation of the form $ \cB_N (t_1,t_2,t_3) = 1/Y(t) ^2\, \cB_N(\hat t_1,\hat t_2, \hat t_3)$ where $\hat t_i= t_i/Y(t)$ and $Y(t)=t_1t_2+t_2 t_3 + t_3 t_1$.  The unknown function $ \cB_N (t_1,t_2,t_3)$ must also be well-defined when $t_i =0$ and the large-$N$ expansion of  \eqref{eq:fullconj} would have to reproduce the terms with half-integer as well as integer powers of $1/N$. Perhaps a recursion relation could be found for $ \cB_N (t_1,t_2,t_3)$ that would relate it to $B_N(t)$ given in \eqref{eq:gsun}, just as we found a recursion relation connecting parts of $\cH_N(\tau)$ to $\cC_N(\tau)$. This would also allow us to compute the {contribution to $\cH_N(\tau)$ that is non-perturbative in $N$, analogous to that found} for $\cC_N(\tau)$ in \cite{Dorigoni:2022cua}.

The identities we found that relate $\cH_N(\tau)$ to $\cC_N(\tau)$ suggest that higher mass derivatives of the partition function $Z_N$ might also be related to lower mass derivatives, or at least certain parts of the large-$N$ expansion as found in this {paper}. From the localisation expression for $Z_N$, it is clear that the perturbative part of the $2p$-th mass derivative of $Z_N$ can be written in terms of products of $p$ zeta functions. While the stress tensor correlator was recently shown to contain single-valued multiple valued zeta functions \cite{Alday:2022xwz,Alday:2023mvu}, it appears that these vanish after taking the integral of the correlator that is related to mass derivatives of $Z_N$\footnote{Similar phenomena were observed in the weak coupling regime~\cite{Wen:2022oky, Brown:2023zbr}.}. It would be interesting to find the modular completions of these products of zeta functions.

The $\tau$-independent terms in the large-$N$ expansions of $\cH_N(\tau)$ and $\cC_N(\tau)$ give two constraints at each orders in the loop expansion of the holographic correlator. At one loop, there was a single contact-term ambiguity, which was fixed using $\cC_N(\tau)$ in \cite{Chester:2019pvm}, and verified using $\cH_N(\tau)$ in this {paper}. At two loops, there are four such ambiguities, so we need two additional constraints. It is possible that derivatives {with respect to the}  squashing parameter $b$ of $Z_N$ on the squashed sphere might give an additional independent constraint. It would be interesting to {find another} source of constraints, so that the 2-loop term can be fixed completely.

The methods developed in this paper to compute the perturbative contributions to $\cH_N(\tau)$ should also apply to other 4d $\mathcal{N}=2$ gauge theories, since the localisation expressions for all 4d $\mathcal{N}=2$ gauge theories include Barnes G-functions that lead to Bessel functions at large $N$. Indeed, our methods were already used to fix the $O(N^0)$ term in the $USp(2N)$ gauge theory considered in \cite{Beccaria:2022kxy,Behan:2023fqq}.\footnote{Our methods were shared with the authors of those papers prior to this publication.} It would be interesting  to study the large-$N$ expansion of mass derivatives of $Z_N$ for other $\cN=2$ gauge theories.

\section*{Acknowledgments} 

DD and MBG are grateful for the hospitality of the Pollica Physics Centre and to conversations with Don Zagier, Kim Klinger--Logan, Ksenia Fedosova and  Boris Pioline, as well as other participants in the programme `New Connections Between Physics and Number Theory’. The work of LFA is supported by the European Research Council (ERC) under the European Union's Horizon
2020 research and innovation programme (grant agreement No 787185). LFA is also supported in part by the STFC grant ST/T000864/1. SMC is supported by a Royal Society University Research Fellowship, URF\textbackslash R\textbackslash 221310. CW is supported by a Royal Society University Research Fellowship, URF\textbackslash R\textbackslash 221015 and a STFC Consolidated Grant, ST\textbackslash T000686\textbackslash 1 ``Amplitudes, strings \& duality". 

\appendix

\section{Matrix model details}
\label{app:details}

In this Appendix, we will discuss some details of the large-$N$ large-$\lambda$ expansion from Section \ref{sec:largeN}. We will first discuss the pole prescription that gives the leading $O(N^2)$ result in \eqref{4mApp2}. We will then discuss how the $O(N^0)$ term in that expression can be computed using a similar but more complicated calculation due to the double sum.

\subsection{$O(N^2)$ at large $\lambda$}

We start with the with the expression for the 4-body contribution in \eqref{I3}, which we repeat here for ease of access:
\es{I3App}{
 I^{(0)}(\lambda) &=\oint\frac{d\sint}{2\pi i}\frac{d\tint}{2\pi i}\, p(\sint,\tint) c(\sint,\tint)\,,\qquad c(\sint,\tint) =\sum^{\infty}_{\ell=1}c_\ell(\sint,\tint)\,,\\
p(\sint,\tint) &= \frac{3 \lambda^{\sint
   +\tint +2}\sin(\pi\sint)\sin(\pi\tint)\Gamma (2 \sint +4)^2
   \Gamma (2 \tint +4)^2 \zeta (2 \sint +3) \zeta (2 \tint +3)}{  2^{4 \sint +4 \tint +3} \pi ^{2 (\sint +\tint +2)}  }\,,\\
c_\ell(\sint,\tint)&=\frac{\ell (\ell-\sint -3) (\ell-\tint -3) \Gamma (\ell-\sint -3)
   \Gamma (\ell-\sint -1) \Gamma (\ell-\tint -3) \Gamma (\ell-\tint
   -1)}{\pi ^2 \Gamma (\ell+\sint +2) \Gamma (\ell+\sint +3)
   \Gamma (\ell+\tint +2) \Gamma (\ell+\tint +3)}\,.
}
We find it convenient to use the following representation for $c(\sint,\tint)$
\begin{align}\label{czeta}
c(\sint,\tint) &=\sum^{\infty}_{q=0} P^{(3 q)}(\sint,\tint)  \zeta (4 \sint +4 \tint +15+2q)\\
&\nn= \frac{\zeta (4 \sint +4 \tint +15)}{\pi ^2} + P^{(3)}(\sint,\tint)  \zeta (4 \sint +4 \tint +17) + \cdots \, , 
\end{align}
where $P^{(3q)}(\sint,\tint)$ are polynomials of degree $3q$, easily computable by expanding $ c_\ell(\sint,\tint)$ for large $\ell$ and performing the sum over $\ell$. This representation is useful, as it makes explicit poles located at $\mbox{Re}(\sint+\tint)<0$ that are easy to miss otherwise. In addition, we will have poles of the type  $\mbox{Re}(\sint),\mbox{Re}(\tint)<0$. We now compute all contributions to $I^{(0)}(\lambda)$ by degree of complexity. 

First there is a contribution arising from the explicit poles of $ p(\sint,\tint)   c_1(\sint,\tint)$ at $\sint=-1$ and $\tint=-1,-2,-5/2,-7/2,-9/2,\cdots$ (and the same with $\sint \leftrightarrow \tint$). This can be computed analytically to all orders in $1/\lambda$ and gives
\begin{equation}
I^{(0)}_0(\lambda) =6 -\frac{16\pi^2}{\lambda}+3 \sum_{n=2}^\infty \frac{(-1)^{n} 4^{n+2} \pi ^{2 n-\frac{7}{2}} \lambda ^{\frac{1}{2}-n} \Gamma \left(n-\frac{5}{2}\right) \Gamma \left(n-\frac{3}{2}\right) \Gamma \left(n+\frac{1}{2}\right) \zeta '(2-2 n)}{\Gamma (n-1) \Gamma (2 n-2)}\,.
\end{equation}

Naively there is a contribution when $\sint=-m,\tint=-n$, with $m,n=2,3,\cdots$. It turns out however, that this contribution is zero, as $c(-m,-n)=0$. This can be seen from the representation for $c(\sint,\tint)$ in terms of zeta functions. For negative integer $\sint,\tint$ the polynomials are such that the series truncate, and we can directly evaluate $ c(\sint,\tint)$ at those points and see that it vanishes. Then there is a contribution from $\sint=-m-1/2,\tint=-n-1/2$, with $m,n=2,3,4,\cdots$. This contribution can be written as
\begin{align}
I^{(0)}_1(\lambda) =3 \sum^{\infty}_{m,n=2} \frac{(-1)^{m+n} 2^{4 m+4 n-1} \pi ^{2 (m+n-1)} \zeta '(2{-}2 m) \lambda ^{-m-n+1} \zeta '(2{-}2 n)}{\Gamma (2 m-2)^2 \Gamma (2 n-2)^2} c(-m{-}\tfrac{1}{2},-n{-}\tfrac{1}{2})\,,
\end{align}
where $c(-m{-}1/2,-n{-}1/2)$ can be easily computed for any pair of integers $m,n$. For  $\sint=-m-1/2,\tint=-n-1/2$ we find
\begin{eqnarray}
c(\sint,\tint) = \frac{\Gamma (-\sint -1) \Gamma (-\sint ) \Gamma (-\tint -1) \Gamma (-\tint )}{6 \pi ^2 \Gamma (\sint +3) \Gamma (\sint +4) \Gamma (\tint +2) \Gamma (\tint +3)} P_\sint(\tint)\,,
\end{eqnarray}
where the polynomials $P_\sint(\tint)$ can be computed for $\sint=-5/2,-7/2,\cdots$
\es{Ps}{
P_{-5/2}(\tint)&=1,\\
P_{-7/2}(\tint)&=\frac{1}{5} \left(-8 \tint ^2-12 \tint +5\right),\\
P_{-9/2}(\tint)&=\frac{1}{105} \left(-128 \tint ^4-128 \tint ^3+272 \tint ^2+152 \tint +105\right),
}
and so on. Lastly, we have the poles at $\sint \! = \! -m$ and $\tint \! = \! -n-1/2$ (and the same with $\sint,\tint$ exchanged). These are slightly subtle because the zeta functions have poles at these locations. We find their contribution is given 
 \begin{align}
 \label{morePoles}
I^{(0)}_2(\lambda) &= \sum^{\infty}_{m=2} \frac{2^{2 m+1} \pi ^{2 m-\frac{5}{2}} \lambda ^{\frac{3}{2}-m} \zeta (3-2 m) \Gamma \left(m-\frac{5}{2}\right) \Gamma \left(m-\frac{3}{2}\right)}{ \Gamma \left(\frac{9}{2}-m\right) \Gamma (m-2) \Gamma (2 m-3)} \nn \\
&  - 3\sum^{\infty}_{m=2}\frac{2^{2 m+2} \pi ^{2 m-\frac{5}{2}} \lambda ^{\frac{1}{2}-m} \zeta (3-2 m) \Gamma \left(m-\frac{3}{2}\right) \Gamma \left(m-\frac{1}{2}\right)}{\Gamma \left(\frac{7}{2}-m\right) \Gamma (m-1) \Gamma (2 m-3)}\\
&  -3\sum^{\infty}_{m=2} \frac{64 (-1)^m \pi ^{2 m-3} \lambda ^{\frac{1}{2}-m} \Gamma \left(m-\frac{5}{2}\right) \Gamma \left(m-\frac{3}{2}\right) \zeta '(2-2 m)}{\Gamma (m-1)^2} \nn \\
&-3\sum^{\infty}_{m,n=2} \frac{4^{m+n+1} \pi ^{2 m+2 n-\frac{9}{2}} \zeta (3-2 m) \lambda ^{-m-n+\frac{3}{2}} \Gamma \left(m+n-\frac{5}{2}\right) \Gamma \left(m+n-\frac{3}{2}\right) \zeta '(2-2 n)}{2\Gamma (2 m-3) \Gamma (2 n-2) \Gamma \left(-m-n+\frac{9}{2}\right) \Gamma (m+n-2)} \,. \nn
\end{align}

Now we discuss poles where $\sint+\tint=negative$, which are apparent from the expression for $c(\sint,\tint) $ as a sum over zeta-functions in \eqref{czeta}. More precisely 
\begin{equation}
c(\sint,\tint)\sim \frac{R_q(\tint)}{\sint-\frac{1}{2} (-(2 q+7)-2 \tint )}\, , \qquad {\rm with} \quad q=0,1,2,\cdots \, ,
\end{equation}
where we will first perform the integral over $\sint$. The residues at those poles can be calculated, and are given by
\begin{equation}
R_q(\tint)=-\frac{4^{-q-1} \Gamma \left(q-\frac{1}{2}\right) \Gamma \left(q+\frac{1}{2}\right) \Gamma \left(q+\frac{3}{2}\right) \Gamma (2 (q+\tint +2))}{\pi ^{7/2} \Gamma (q+1) \Gamma (2 \tint +4)}\,.
\end{equation}
We can now perform the integral over $\sint$ for each $q$, and obtain
\begin{align}
I^{(0)}_{3}(\lambda) & = \sum^{\infty}_{q=0} \lambda ^{-q-\frac{3}{2}} I_{3,q}^{(0)}\,,\qquad{\rm where}\qquad I_{3,q}^{(0)}\equiv   \oint \frac{d\tint}{2\pi i} Q_q(\tint) \, , 
\end{align}
with 
\begin{align}
&Q_q(\tint)=\\
&\nn 3\frac{(-1)^{q+1} 4^{q+4}\Gamma \left(q{-}\frac{1}{2}\right) \Gamma \left(q{+}\frac{1}{2}\right) \Gamma \left(q{+}\frac{3}{2}\right) \Gamma (2 \tint +4) \zeta (2 \tint +3) \Gamma (-2 q-2 \tint -3) \zeta (-2 (q+\tint +2))}{ \pi ^{-\frac{1}{2}-2q}  \Gamma (q+1)} \,.
\end{align}
There is a subtle question of which poles to include in the integrals for $I^{(0)}_{3}(\lambda)$. This is akin to the precise choice of the contour. It turns out we need to choose all poles for negative $\tint$. Let's consider the leading order contribution, arising from $q=0$, as this already has the ingredients we will encounter later on:
\begin{eqnarray}
I^{(0)}_{3,0} = 3\pi^2 \oint \frac{d\tint}{2\pi i} 2^8 \Gamma (-2 \tint -3) \Gamma (2 \tint +4) \zeta (-2 (\tint +2)) \zeta (2 \tint +3) \, , 
\end{eqnarray}
summing over all poles with $\mbox{Re}(\tint)<0$ we can see that the poles in the region $-4<\tint<0$  cancel each other, and we obtain the following sum
\begin{eqnarray}
I^{(0)}_{3,0}= 3\pi^2 2^7 \sum_{n=4}^\infty \zeta (3-2 n) \zeta (2 n-4)\,.
\end{eqnarray}
This is a divergent sum, and we should regularise it. Using the integral representation
\begin{eqnarray}
\zeta(n) = \frac{1}{\Gamma(n)} \int_0^\infty \frac{x^{n-1}}{e^x-1} dx \, , 
\end{eqnarray}
which is valid for $n>1$, together with the reflection relation, we can obtain the following expression
\begin{eqnarray}
I^{(0)}_{3,0}= -3\sum^{\infty}_{n=4} \int_0^\infty dx dy \frac{2^{10-2 n} \pi ^{4-2 n} x^{2 n-3} y^{2 n-5} \cos (\pi  n)}{\left(e^x-1\right) \left(e^y-1\right) \Gamma (2 n-4)}\,.
\end{eqnarray}
We can now exchange the sum and the integral, and perform the sum over $n$ to find
\begin{eqnarray}
I^{(0)}_{3,0}=48 \int_0^\infty dx dy\frac{x^2 \left(2 \pi  \sin \left(\frac{x y}{2 \pi }\right)-x y\right)}{\pi ^2 \left(e^x-1\right) \left(e^y-1\right)}\,,
\end{eqnarray}
which is a convergent double integral that yields
\begin{eqnarray}
I^{(0)}_{3,0} =-{96 \zeta (3)}-\frac{8\pi ^4}{15}+16\pi^2\,.
\end{eqnarray}
One may be worried about the regularisation procedure. However, the integral over $\tint$ can also be computed numerically, choosing a contour slightly to the left of the imaginary axis, and the result agrees perfectly with our analytic result. At higher orders we can do exactly the same, but the sums involved are a bit harder:
\begin{equation}
I^{(0)}_{3,q} =3 \sum^{\infty}_{n=4+q} \frac{(-1)^{q+1}2^{2 q+7}  \Gamma \left(q{-}\frac{1}{2}\right) \Gamma \left(q{+}\frac{1}{2}\right) \Gamma \left(q{+}\frac{3}{2}\right)  \Gamma (2 n{-}2 q{-}3)}{\pi ^{-1/2} \Gamma (2 n-3) \Gamma (q+1)}\zeta (3{-}2 n)\zeta (2 n{-}2 q{-}4)\,.
\end{equation}
Introducing an integral representation for the zeta functions and performing the sum over $n$ we obtain
\begin{align}
I^{(0)}_{3,q} &=-48\pi^2\int_0^\infty dx dy \frac{y^3 x^{2 q+5} \Gamma \left(q-\frac{1}{2}\right) \Gamma \left(q+\frac{1}{2}\right) \Gamma \left(q+\frac{3}{2}\right) H_q(x,y)}{(2 \pi )^{2 q+9}  \left(e^x-1\right) \left(e^y-1\right) \Gamma (q+1)},\\
H_q(x,y)&= x^2 y^2 \, _1\tilde{F}_2\left(2;q+\frac{7}{2},q+4;-\frac{x^2 y^2}{16 \pi ^2}\right)-32 \pi ^2 \, _1\tilde{F}_2\left(1;q+\frac{5}{2},q+3;-\frac{x^2 y^2}{16 \pi ^2}\right)\,. \nonumber
\end{align}
For any integer value of $q$ the regularised hypergeometric functions reduce to something simpler. The integrals over $x$ are always of the form
\begin{align}
\int_0^\infty dx \frac{x^a}{e^x-1} &= \Gamma (a+1) \zeta (a+1)\,,\\
\int_0^\infty dx \frac{x \cos \left(\frac{x y}{2 \pi }\right)}{e^x-1}&= \frac{1}{2} \pi ^2 \left(\frac{4}{y^2}-\text{csch}^2\left(\frac{y}{2}\right)\right), \\
\int_0^\infty dx \frac{x^2 \sin \left(\frac{x y}{2 \pi }\right)}{e^x-1}&= \pi ^3 \left(\frac{8}{y^3}-\frac{1}{2} \sinh (y) \text{csch}^4\left(\frac{y}{2}\right)\right)\,.
\end{align}
For every fixed $q$, this allows to perform the integral over $x$. The integral over $y$ can also be performed, using\footnote{In practise, the simplest way to perform the integrals over $y$ is to multiply by $y^\epsilon$, use the expressions below and then take $\epsilon \to 0$ at the end.} 
\begin{align}
\int_0^\infty dy \frac{y^a \text{csch}^2\left(\frac{y}{2}\right)}{e^y-1} &= \Gamma(a+1) (2 \zeta (a-1)-2 \zeta (a))\,,\\
\int_0^\infty dy \frac{y^a \sinh (y) \text{csch}^4\left(\frac{y}{2}\right)}{e^y-1} &= \frac{4\Gamma(a+1)}{3}(2 \zeta (a-2)-3 \zeta (a-1)+\zeta (a))\,.
\end{align}
With these results we can compute the integrals for any given $q$ and we can actually write down the contributions analytically. We find a structure which resembles very much $I^{(0)}_2(\lambda)$. More precisely we get
\begin{align}
\label{I30lam}
&I^{(0)}_3(\lambda) =-\sum^{\infty}_{m=2} \frac{2^{2 m+1} \pi ^{2 m-\frac{5}{2}} \lambda ^{\frac{3}{2}-m} \zeta (3-2 m) \Gamma \left(m-\frac{5}{2}\right) \Gamma \left(m-\frac{3}{2}\right)}{\Gamma \left(\frac{9}{2}-m\right) \Gamma (m-2) \Gamma (2 m-3)} \nn \\
&  -3 \sum^{\infty}_{m=2}  \kappa_1(m) \frac{2^{2 m+2}  \pi ^{2 m-\frac{5}{2}} \lambda ^{\frac{1}{2}-m} \zeta (3-2 m) \Gamma \left(m-\frac{3}{2}\right) \Gamma \left(m-\frac{1}{2}\right)}{\Gamma \left(\frac{7}{2}-m\right) \Gamma (m-1) \Gamma (2 m-3)}\\
&  -3\sum^{\infty}_{m=2}  \kappa_2(m)  \frac{64 (-1)^m \pi ^{2 m-3} \lambda ^{\frac{1}{2}-m} \Gamma \left(m-\frac{5}{2}\right) \Gamma \left(m-\frac{3}{2}\right) \zeta '(2-2 m)}{\Gamma (m-1)^2} \nn \\
&-3\!\!\sum^{\infty}_{m,n=2} \!\! \kappa_1(m) \frac{4^{m+n+1} \pi ^{2 m+2 n-\frac{9}{2}} \zeta (3-2 m) \lambda ^{-m-n+\frac{3}{2}} \Gamma \left(m+n-\frac{5}{2}\right) \Gamma \left(m+n-\frac{3}{2}\right) \zeta '(2-2 n)}{2\Gamma (2 m-3) \Gamma (2 n-2) \Gamma \left(-m-n+\frac{9}{2}\right) \Gamma (m+n-2)} \, , \nn
\end{align}
with
\es{kappa}{
\kappa_1(m) =2\delta_{m,2}-1\,,\qquad 
\kappa_2(m)= -\frac{(m-3)(2m-1)}{3} \,, \\
}
Note that we can combine the last two contributions into
\begin{align}
I^{(0)}_2(\lambda) +I^{(0)}_3(\lambda) &= \frac{32}{ \lambda ^{3/2}} - \sum^{\infty}_{m=2} \frac{64 (-1)^m m(7-2 m) \pi ^{2 m-3} \lambda ^{\frac{1}{2}-m}  \Gamma \left(m-\frac{5}{2}\right) \Gamma \left(m-\frac{3}{2}\right) \zeta '(2-2 m)}{ \Gamma (m-1)^2} \nonumber \\
 & + \sum^{\infty}_{m=2} \frac{4^{m+2} \pi ^{2 m-\frac{1}{2}} \lambda ^{-m-\frac{1}{2}} \Gamma \left(m-\frac{1}{2}\right) \Gamma \left(m+\frac{1}{2}\right) \zeta '(2-2 m)}{ \Gamma \left(\frac{5}{2}-m\right) \Gamma (m) \Gamma (2 m-2)}\,.
\end{align}
Finally, the derivative of Riemann zeta function can be simplified using the identity
\begin{equation}
\zeta'(-2m) = \frac{(-1)^m (2m)!}{2\pi^{2m}} \zeta(2m+1)\,,
\end{equation}
valid for any $m$ strictly positive integer.
We can then put all ingredients together and compute $I^{(0)}(\lambda)$ to any desired order. The lowest few orders match \eqref{secondTermNFac}, while higher orders are given in the attached \texttt{Mathematica} file. 

\subsection{$O(N^0)$ at large $\lambda$}

Recall that there are two contributions to the 4-body term in \eqref{Js}: those that factorise in $\omega_1,\omega_2$, and those that do not. The factorizable terms $\mathcal{J}_0^\text{fac}(\omega_1,\omega_2)$ can be computed at large $\lambda$ just like the 2-body terms by writing them in Mellin space using \eqref{mbbessel}, performing the $\omega_1,\omega_2$ integrals using \eqref{id}, and closing each contour to the left. We obtain
\es{facEasy}{
I^{(2)}_\text{fac}(\lambda)=-\frac{13 \sqrt{\lambda
   }}{6}+\frac{55}{12}+\frac{3 }{8\sqrt{\lambda}}-\frac{13 \zeta (3)+6}{4 \lambda }
  +\frac{273}{64} \frac{\zeta
   (3)}{\lambda^{3/2} } +O(\lambda^{-2})\,.
}
For the non-factorizable terms $\mathcal{J}_0^\text{fac}(\omega_1,\omega_2)$, the first two lines in \eqref{Js} take a similar form as the $O(N^2)$ terms in the previous section, i.e. they have $(\omega_2^2-\omega_1^2)$ in the denominator, so they can be evaluated using similar methods to get an ``easy'' contribution
\es{facMid}{
I^{(2)}_{easy}(\lambda)=\frac{\sqrt{\lambda }}{6}+\frac{17}{12} +\left(-\frac{31}{8}-\frac{\pi ^2}{6}\right) {\frac{1}{\sqrt{\lambda}
   }}+\frac{\zeta
   (3)+6}{4 \lambda }+
   \left(\frac{51 \zeta (3)}{64}+\frac{5 \pi ^2}{8}\right)\frac{1}{\lambda^{3/2}}+O(\lambda^{-2})\,.
}
The other terms in $\mathcal{J}_0^\text{fac}(\omega_1,\omega_2)$ have $(\omega_2^2-\omega_1^2)^2$ in the denominator, so they require a slightly more complicated approach. As with the other non-factorizable terms, we use the identity \eqref{BesKern} to write these ``hard'' terms as 
\begin{equation}
I^{2}_{hard}(\lambda) = \oint \frac{d\sint}{2\pi i}  \frac{d\tint}{2\pi i} p(\sint,\tint) c(\sint,\tint),~~~c(\sint,\tint) = \sum^\infty_{\substack{k=1 \\ \ell=0}} c_{k,\ell}(\sint,\tint)
\end{equation}
with 
\begin{equation}
p(\sint,\tint)=3\frac{\sin (\pi  \sint ) \sin (\pi  \tint )  \Gamma (2 \sint +4) \Gamma (2 \sint +6)\Gamma (2 \tint +4)^2 \zeta (2 \sint +5) \zeta (2 \tint +3) \lambda ^{\sint +\tint +3}}{2^{4 \sint +4 \tint +7}\pi ^{2 (\sint +\tint +4)}}
\end{equation}
and
\begin{equation}
c_{k,\ell}(\sint,\tint)=\frac{k (2 \ell+1) \Gamma (k-\ell-\sint -2) \Gamma (k+\ell-\sint -1) \Gamma (k-\ell-\tint -2) \Gamma (k+\ell-\tint -1)}{\Gamma (k-\ell+\sint +2) \Gamma (k+\ell+\sint +3) \Gamma (k-\ell+\tint +2) \Gamma (k+\ell+\tint +3)}\,.
\end{equation}
We can apply the method described above by changing variables, and writing $c(\sint,\tint) = \sum^\infty_{p=1} c_{p}(\sint,\tint)$ where each $c_{p}(\sint,\tint)$ is given by a finite sum
\begin{equation}
c_{p}(\sint,\tint)=\sum_{q=1}^p \frac{q (2 p-2 q+1) \Gamma (p-\sint -1) \Gamma (p-\tint -1) \Gamma (-p+2 q-\sint -2) \Gamma (-p+2 q-\tint -2)}{\Gamma (p+\sint +3) \Gamma (p+\tint +3) \Gamma (-p+2 q+\sint +2) \Gamma (-p+2 q+\tint +2)}\, .
\end{equation}
In this form it can be shown that $c_{p}(\sint,\tint)$ satisfies a second order homogeneous recursion relation 

\begin{equation}
f_0(p) c_{p}(\sint,\tint)+ f_1(p) c_{p+1}(\sint,\tint)+f_2(p) c_{p+2}(\sint,\tint)=0\,,
\end{equation}
with
\begin{align}
f_0(p)&=(p+2)^2 (p-\sint -1) (p-\sint )^2 (p-\tint -1) (p-\tint )^2\,, \\
f_1(p)&=-(p-\sint ) (p+\sint +3) (p-\tint ) (p+\tint +3)  \\
 \times  & \left   [\left(2 p^2+6 p+5\right) \sint  \tint +\left(5 p^2+15 p+11\right) (\sint +\tint )+2 p^4+12 p^3+36 p^2+54 p+29\right]\,, \nn \\
f_2(p)&=(p+1)^2 (p+\sint +3)^2 (p+\sint +4) (p+\tint +3)^2 (p+\tint +4)\,.
\end{align}
Supplemented with the leading order terms in the large $p$ expansion, this recursion relation allows us to compute arbitrarily high orders. The final answer takes the form
\begin{align}
c_{p}(\sint,\tint)&= \frac{1}{(2 \sint +2 \tint +5) (2 \sint +2 \tint +7)} \frac{1}{p^{4 \sint +4 \tint +13}} \left(1-\frac{2 \sint +2 \tint +\frac{13}{2}}{p}+ \cdots \right) \\
& +\frac{\pi ^{3/2} 4^{\sint +\tint +2} \csc (\pi  \sint ) \csc (\pi  \tint ) \Gamma \left(\sint +\tint +\frac{7}{2}\right)}{\Gamma (2 \sint +4) \Gamma (2 \tint +4) \Gamma (\sint +\tint +4)} \frac{1}{p^{2(3+\sint+\tint)}} \left(1- \frac{\sint +\tint +3}{p} + \cdots\right)\,, \nn
\end{align}
 with two distinct type of terms. Summing over $p$ term by term we obtain a representation analogous to \eqref{czeta}. The rest of the computation proceeds as above and yields
 \es{hard2}{
 I^{2}_{hard}(\lambda) =\left(4-\frac{4 \pi ^2}{15}\right) \sqrt{\lambda }+\frac{ 24 \zeta (3)-83}{10}+  \frac{21+\pi ^2}{6\sqrt{\lambda}}+\frac{3}{16}
   \frac{\left(12 \zeta (3)+\pi
   ^2\right)}{\lambda^{3/2}}+O(\lambda^{-2})\,.
 } 
We then sum \eqref{facEasy}, \eqref{facMid}, and \eqref{hard2} along with the 2-body term \eqref{Ieasy} and the $-12\zeta(3)$ in \eqref{4m} to get the final answer in \eqref{4mApp2}.

 \section{Generalised Eisenstein series}
 \label{sec:genais}
 
 A Generalised Eisenstein Series satisfies the inhomogeneous Laplace eigenvalue equation
   \bea
(\Delta_\stau -s(s-1) )\, \cE(s,s_1,s_2; \stau) =  E (s_1;\stau)  
E(s_2;\stau) \,. 
 \label{eq:lapfinalApp}
 \eea
 Whereas non-holomorphic Eisenstein series arise as the coefficients of the $R^4$ and $d^4R^4$ interactions  in the low-energy expansion of four-graviton amplitude in type IIB superstring theory, the coefficient of the $d^6R^4$ interaction is the generalised non-holomorphic Eisenstein series $\cE(4,\threeh,\threeh; \stau)$ \cite{Green:2005ba}.   More general examples with half-integral indices in the source term arise in the  low-energy expansion of two-loop eleven-dimensional  supergravity amplitudes compactified on a two-torus \cite{Green:1999pu}. This is the prototype for this general type of modular functions that arise in $\cH_N^i(\tau \btau)$.  For generic eigenvalue $s$, we define the generalised Eisenstein series as the unique modular invariant solution to \eqref{eq:lapfinalApp} satisfying the boundary condition that the coefficient of the homogeneous solution $\tau_2^{s}$ vanishes.

 Within string theory contexts, the eigenvalue spectrum of $\cE(s,s_1,s_2;\stau)$ is sensitive to whether  $s_1$, $s_2$ are integer or half-integer.
It has been argued \cite{Dorigoni:2021jfr} that when $s_1, s_2 \in \NN$  the spectrum of $s$ is restricted to
$s\in\mathrm{Spec}(s_1,s_2) = \{ |s_1-s_2|+2,\,   |s_1-s_2|+4, \dots,s_1+s_2-4, \, s_1+s_2-2\}$.  However, for the generalised Eisenstein series  with $s_1,s_2\in \NN+{1\over 2}$  arising 
as coefficients of the   $1/N^r$ terms in the large-$N$ expansion of the integrated correlator $\cH_N$ \cite{Chester:2020vyz}, the spectrum is found to be restricted to $s=\{s_1+s_2+1, s_1+s_2 +3, \dots, 3r+1 \}$.  This is discussed further in the main text. 
 
The generalised Eisenstein series $\cE(s,s_1,s_2;\tau)$ can be decomposed in Fourier modes in the form
\begin{equation}
\cE(s,s_1,s_2;\tau) = \sum_{n,m= 0 }^\infty \cE^{(n,m)}(s,s_1,s_2;\tau_2) q^n \bar{q}^m\,,
\end{equation}
 where $q=e^{2 \pi i \stau}$ and $\bar q = e^{-2\pi i\bar \stau}$.  The non-zero modes represent the contributions of instantons and anti-instantons,  with instanton number $k=n-m$, where $k>0$ for instantons and $k<0$ for anti-instantons.  Terms with $k=0$ get purely perturbative contributions from the $m=n=0$ term and instanton/anti-instanton contributions from terms with $m=n >0$.
 
 The Laurent polynomial of $ \cE(s,s_1,s_2;\tau_2)$  (i.e. the $\cE^{(0,0)}$  term) is given by
\begin{align}
\label{genEisPert}
&\cE^{(0,0)}(s,s_1,s_2;\tau_2) = \nn \\ 
& \frac{4 \pi^{-2s_1-2s_2} \zeta(2s_1) \zeta(2s_2)}{\pi^{s_1+s_2}(s_1+s_2-s)(s_1+s_2+s-1)} (\pi \tau_2)^{s_1+s_2} + \frac{4 {\pi^{-{1\over 2}-2s_2}}\Gamma(s_1-\half) \zeta(2s_1-1)\zeta(2s_2)}{(s_2-s_1+s)(s_2-s_1-s+1) \Gamma(s_1)} (\pi \tau_2)^{1-s_1+s_2} \nn \\
 +&  \frac{4 \pi^{-{1\over 2}-2s_1}\Gamma(s_2-\half) \zeta(2s_2-1)\zeta(2s_1)}{(s_1-s_2+s)(s_1-s_2-s+1)
 \Gamma(s_2)} (\pi \tau_2)^{1-s_2+s_1}
\\
+ & \frac{4 \Gamma(s_1-\half)\Gamma(s_2-\half) \zeta(2s_1-1)\zeta(2s_2-1)}{\pi (s_1+s_2-s-1)(s_1+s_2+s-2) \Gamma(s_1)\Gamma(s_2)} (\pi \tau_2)^{2-s_1-s_2}
+ \beta(s,s_1,s_2) \,   (\pi \tau_2)^{1-s} \,, \nn
\end{align}
where the first four terms can be obtained by matching powers of $\tau_2$ on the left-hand and right-hand sides of \eqref{eq:lapfinalApp} and the last term satisfies the homogeneous equation and  its coefficient is given by \cite{Green:2005ba,Green:2008bf}\footnote{In these references, this coefficient is determined by projecting the Laplace equation  \eqref{eq:laplacewsource} on $E(s;\stau)$. Alternatively, it may be obtained from a Poincar\'e series representation as in \cite{Dorigoni:2021jfr}.}

 \begin{equation}
 \beta(s,s_1,s_2)  = {4 \pi^{s-1}\over
  \Gamma(s_{1})\Gamma(s_{2})}\,
{\zeta^*(s-s_{1}-s_{2}+1)\zeta^*(s+s_{1}-s_{2})
\zeta^*(s-s_{1}+s_{2})\zeta^*(s+s_{1}+s_{2}-1)\over (1-2s)\,\zeta^*(2s)}\,,
\end{equation}
 with $\zeta^*(s)=\zeta(s)\Gamma(s/2)/\pi^{s/2}$.  
 
 In addition to the Laurent expansion, the $k=0$ sector contains contributions from instanton/anti-instanton pairs, $(n,n)$ with $n>0$, which have the form
\begin{equation}
\cE^{(n,n)}(s,s_1,s_2;\tau_2)  =\frac{ n^{s_1 + s_2 - 2} \sigma_{1-2s_1}(n)\sigma_{1-2s_2}(n) }{\Gamma(s_1)\Gamma(s_2)}\Phi_{s,s_1,s_2}( 4\pi n \tau_2)\,.
\label{eq:Fqq}
\end{equation}
When $s_1,s_2\in \NN$ the expression $\Phi_{s, s_1,s_2}( 4\pi n\tau_2)$ is a polynomial  in inverse powers of $\tau_2$ of degree $s_2+s_1-2$.
The first two perturbative orders in the $(n,n)$ sector do not depend on the eigenvalue $s$ and have the form
\begin{align}
\Phi_{s,s_1,s_2}( 4\pi n \tau_2)  =  \frac{8}{(4\pi n \tau_2)^2}+ \frac{8[s_1(s_1-1)+s_2(s_2-1) -4]}{(4\pi n \tau_2)^3}+O(\tau_2^{-4})\, .
\end{align}
 Higher-order coefficients do depend on $s$ and can be computed from the differential equation \eqref{eq:lapfinalApp} as performed in \cite{Chester:2020vyz, Fedosova:2022zrb} or via resurgence analysis methods \cite{Dorigoni:2022bcx,Dorigoni:2023nhc}. The contribution of the general $(n,m)$ sector with $n,m\neq 0$ is given by
\begin{equation}
\cE^{(n,m)}(s,s_1,s_2;\tau_2)= \frac{ n^{s_1 -1}m^{s_2-1} \sigma_{1-2s_1}(n)\sigma_{1-2s_2}(m) }{\Gamma(s_1)\Gamma(s_2)}\Big(
\frac{1}{4 nm\, (\pi \tau_2)^2} +O(\tau_2^{-3})\Big) + (n\leftrightarrow m) \,.\label{eq:Fqqbn1n2}
\end{equation}
{These terms correspond to non-perturbative contributions in the topological sector with instanton number $n-m$.}

{For the purely instantonic sector, i.e.~the $(n,0)$ term, and purely anti-instantonic sector, i.e.~the $(0,m)$ term, we need to be more careful since in these sectors we  now have the freedom of adding homogeneous solutions to the differential equation \eqref{eq:lapfinalApp}.}
An algorithm for constructing a particular solution, $\cE_{{\rm p}}(s,s_1,s_2;\tau)$, to \eqref{eq:lapfinalApp} was proposed in \cite{Chester:2020vyz}, and later generalised in \cite{Fedosova:2022zrb}.
As discussed in the main text, whenever the eigenvalue $s$ is such that the vector space, $\mathcal{S}_{2s}$, of holomorphic cusp forms with weight $2s$ is non-trivial, this particular solution $\cE_{{\rm p}}(s,s_1,s_2;\tau)$ does not correspond to a modular invariant solution to \eqref{eq:lapfinalApp}, i.e. the particular solution does not correspond to the generalised Eisenstein series.

To construct the generalised Eisenstein series $\cE(s,s_1,s_2;\tau)$, a certain homogeneous solution has to be added  to the particular solution, so that the complete solution is a modular function. Since we have fixed the boundary condition in the perturbative, i.e. the $(0,0)$ sector, this is only relevant for the instantonic (and anti-instantoic) pieces, namely we can only add an homogeneous solution to $\cE_{\rm p}^{(n,0)}(s,s_1,s_2;\tau_2)$ (and $\cE_{\rm p}^{(0,m)}(s,s_1,s_2;\tau_2)$). 

Based on  \cite{Dorigoni:2021jfr,Dorigoni:2021ngn}, we can use the novel results \cite{FKRinprogress} to show that even in the case of present interest the homogeneous solution that must be added is given by
\ie
\lambda_{\Delta}(s,s_1,s_2)  \cdot H_{\Delta}(\tau) \, ,
\fe
where  the coefficient $\lambda_{\Delta}(s,s_1,s_2)$ is presented in \eqref{eq:lvalue}. The function $H_{\Delta}(\tau)$ takes the form, 
\begin{equation}
H_{\Delta}(\tau)\equiv \sum_{n=1}^\infty  \frac{a_\Delta(n)}{n^s} \sqrt{n\tau_2 }  K_{s-\frac{1}{2}}(2 \pi n \tau_2)\big( e^{2\pi i n \tau_1}+e^{-2\pi i n \tau_1}\big)\, ,
\end{equation}
and satisfies the homogeneous equation 
\begin{equation}
(\Delta_\tau-s(s-1)) H_{\Delta} (\tau)= 0\, . 
\end{equation}
Note the  important fact that $H_{\Delta} (\tau)$ alone is not modular invariant.  However, the addition of a suitable multiple of $ H_{\Delta}(\tau)$ to the particular solution $\cE_{{\rm p}}(s,s_1,s_2;\tau)$  is crucial for obtaining a modular invariant solution to \eqref{eq:lapfinalApp}, i.e. for constructing the generalised Eisenstein series
\begin{equation}
 \cE(s,s_1,s_2;\tau)=  \cE_{{\rm p}}(s,s_1,s_2;\tau) + \sum_{\Delta\in \cS_{2s}} \lambda_{\Delta}(s,s_1,s_2) H_{\Delta}(\tau)\,.
\end{equation}

\section{Properties of lattice sums}
\label{sec:geneis}

In general there is no known lattice integral representation for an individual generalised Eisenstein series.
However, we shall now describe the construction of a lattice representation  of the special linear combinations of generalised Eisenstein series which are relevant for the integrated correlator under discussion
(see \cite{Green:1999pu,Green:2005ba,Green:2008bf,DHoker:2018mys,Bossard:2020xod} for further details).  This gives a four-dimensional lattice representation that  generalises the two-dimensional lattice representation of a non-holomorphic Eisenstein series.

\subsection{A family of lattice sum integrals}
Following \cite{Green:2005ba}, we here consider a class of lattice sum integrals constructed from the special non-holomorphic functions $A_{i,j}(\ttau)$, called \textit{modular local polynomials}, introduced in section~\ref{sec:lattice}  and  whose precise definition and properties will be reviewed in appendix \ref{eq:aijdef}. 
We will consider the lattice sum integral in the form\footnote{In earlier lieterature, such as \cite{Green:2005ba} the variables $t_1$, $t_2$ $t_3$ corresponded to `inverse Schwinger parameters' $t_i = \hat{t}_i/(\hat{t}_1\hat{t}_2+\hat{t}_1 \hat{t}_3+\hat{t}_2\hat{t}_3)$, where $\hat{t}_i$ ($i=1,2,3$)  are the Schwinger parameters for a two-loop vacuum diagram.}
\begin{align}  
\cE^w_{i,j}(\stau) =   \sum_{\underset {p_1+p_2+p_3=0} {p_1,p_2, p_3\ne 0}} \int_0^\infty d^3 t\, [V(t)]^{w-3}  \, A_{i,j}(\ttau(t))\,  \exp\Big( - \frac{\pi}{\tau_2}\sum_{i=1}^3 t_i |p_i|^2  \Big)\,,
\label{eq:defEwijApp}
\end{align}
where the parameters $\rho(t)$ and $V(t)$ are defined in \eqref{eq:rhodef}.  The $p_i$ are discrete loop momenta of a two-loop Feynman diagram on a two-torus of complex structure $\tau$ and are defined by 
\begin{equation}
p_i =m_i+ n_i \tau\,,
\label{eq:pdef}
\end{equation}
with $m_i,n_i\in \Z$.

The analysis of the inhomogeneous Laplace equation satisfied by $\cE^w_{i,j}(\stau)$ will demonstrate that the parameter $w\in \mathbb{R}^+$ corresponds to the total transcendental weight of the sources, i.e. the sum of the indices of the bilinears in Eisenstein series which appear as source terms.
 
 In  \cite{Green:2005ba} a slightly different lattice sum was considered which gave rise to ultraviolet divergences arising at the
boundary of the integration region from particular $SL(2,\ZZ)$ orbits of the lattice sums variables. With the present constraints on the lattice sum, the integral \eqref{eq:defEwijApp} is completely well-defined since it is exponentially convergent at all boundaries of the domain of integration.

 \subsection{Modular local polynomials: definition and properties}
  \label{eq:aijdef}

We now present the definition and clarify the most important properties of the non-holomorphic functions $A_{i,j}(\ttau)$, which feature prominently in the formula \eqref{eq:defEwijApp} for the modular invariant functions under consideration.

The function  $A_{1,0}(\ttau)$ arose  in the construction of the $d^6R^4$ coefficient $\cE(4;\frac{3}{2},\frac{3}{2};\tau\btau)$  in  \cite{Green:2005ba}, and a more general discussion in  \cite{Green:2008bf} presented the general procedure for constructing particular combinations of $\cE(s,s_1,s_2;\tau\btau)$ once the  $A_{i,j}(\ttau)$ are known. 
 The following  general construction of $A_{i,j}(\ttau)$ was given by Don  Zagier  (in unpublished notes that are reproduced in  \cite{DHoker:2018mys} with more details). 

The result of this analysis will be that $A_{i,j}(\ttau)$  are  Laurent polynomials in $\ttau_2$ with coefficients which are polynomial in $\ttau_1$, and inside the domain $\ttau \in \mathcal{H}\backslash \Gamma_0(2)$ they satisfy the homogeneous Laplace equation \eqref{DeltaOm},
\begin{equation}
\label{DeltaA}
\left[\Delta_\ttau   - s(s-1) \right] \, A_{i,j}  (\ttau)=0\, , 
\end{equation}
where the eigenvalue $s=3i+j+1$ (for example, $s=4$ for  $A_{1,0}$).

The first step in constructing these Laurent series is to  define the complex combinations
\begin{eqnarray}
u= \ttau^2(1- \ttau)^2\, , \qquad v=\ttau^2-\ttau+1 \,.
\label{eq:uvdef}
\end{eqnarray}
The  functions $A_{i,j}(\ttau)$   are given by 
\begin{eqnarray}
\label{defAij}
A_{i,j}(\ttau) = D_{-2n}^{(n)} ( u^i v^j) \, , \hskip 1in n=s-1 = 3 i +j \quad \hbox{ with } \quad i,j \geq 0 \, . 
\end{eqnarray}
 The polynomials $u^i v^j$, holomorphic in $\ttau$, can be understood precisely as the modular local polynomials of modular weight $-2n$ discussed in \cite{Bringmann}.
Here  $D_{-2n}^{(n)}$ is the iterated derivative operator 
\be
\label{eq:Dn}
D_{-2n}^{(n)} = {(-2i)^n n! \over (2n)!} D_{-2} \circ D_{-4} \circ \cdots \circ D_{-2n+2} \circ D_{-2n}\, , 
\ee
where the differential operator $D_k = \p_\ttau + k/(\ttau-\bar \ttau)$
satisfies $\Delta_{k+2} \cdot D_k - D_{k} \Delta_k = -k D_k$, where $\Delta_k=4 D_{k-2} \, \ttau_2^2 \p_{\bar \ttau}$ is the laplacian acting on weight $k$ modular forms. 
Explicitly, the operator $D_{-2n}^{(n)} $ can be written as
\be
D_{-2n}^{(n)} = {(-2i)^n n! \over (2n)!} 
\sum_{m=0}^n \begin{pmatrix} n \cr m \end{pmatrix}\frac{(-n-m)_m}{(\ttau -\bar \ttau )^m}\, \frac{\partial^{n-m}}{\partial \ttau ^{n-m}} \, ,
\ee
where $(x)_m$ is the Pochhammer symbol.

  The functions $A_{i,j}(\ttau)$ constructed in this manner satisfy the Laplace equation \eqref{DeltaA} in the interior of 
the  fundamental domain $\mathcal{H}\backslash \Gamma_0(2)$.  Furthermore they are Laurent expansions that have  the following form,
\be
\label{Aijexp}
A_{i,j}(\ttau) =\sum_{k=0}^{2i+j} A_{i,j}^{(k)}(\ttau_1)\, \ttau_2^{i+j-2k} \, ,
\ee
where $A_{i,j}^{(k)}(\ttau_1) $ is a polynomial of degree  $k$ in $\ttau_1(1-\ttau_1)$. Since it is invariant under $\ttau_1\mapsto 1-\ttau_1$, it may be expressed as a linear combination of Bernoulli polynomials $B_{2k}(\ttau_1)$ of even index.   It is often clarifying to re-express  $\ttau_1, \ttau_2$ in terms 
of $t_1,t_2,t_3$.  The function $A_{i,j}$ is  then by construction a homogenous function of the $t_i$'s, invariant under permutations.

\subsection*{Examples of $A_{i,j}$} 

From \eqref{defAij}, it is easy to generate expressions for $A_{i,j}$ with very high eigenvalues $s=3i+j+1$. Here we display the $A_{i,j}$ that  appear in the main body of the paper --  in particular  in \eqref{eq:Eijw1}-\eqref{eq:Eijw3}.
For convenience,  we express $A_{i,j}$ in terms of the $t$-variables using \eqref{eq:rhodef}.  Since $A_{i,j}(\ttau(t))$ is a symmetric function of $t_i$ we shall make use of a basis $(\sigma_1,\sigma_2,\sigma_3)$ of symmetric polynomials defined by
\begin{equation}
\sigma_1 \equiv t_1 +t_2 +t_3 \,, \qquad \sigma_2 \equiv Y(t) = t_1 t_2 +t_1 t_3 +t_2 t_3 \,,\qquad \sigma_3 \equiv t_1 t_2 t_3\,.\label{eq:symVar}
\end{equation}
 
Below we list examples of $A_{i,j}$ that are relevant for $\cH_N^{i}(\tau)$ up to order $1/N^3$,
\allowdisplaybreaks{
\begin{eqnarray}
\underline{s=4}:&&
  \sigma_2^{\threeh}A_{1,0} = \frac{1}{5} \sigma_1 \sigma_2-  \sigma_3\,  \\
\underline{s=5}:&&\,
 \sigma_2^2 A_{1,1} = \frac{1}{7} \sigma_1^2 \sigma_2+\frac{2}{35}\sigma_2^2 - \sigma_1 \sigma_3\,,\\
\underline{s=6}:  &&
 \sigma_2^{\fiveh} A_{1,2}=\frac{1}{9}\sigma_1^3 \sigma_2 +\frac{1}{21}\sigma_1 \sigma_2^2 -\sigma_1^2\sigma_3 +\frac{1}{3}\sigma_2 \sigma_3\,,
 \\* &&  \sigma_2^{\fiveh}  A_{0,5}=\sigma_1^5 -\frac{10}{3}\sigma_1^3\sigma_2 +\frac{15}{7}\sigma_1 \sigma_2^2 \,,\\
\underline{s=7}:&&
\,\sigma_2^3 A_{2,0} = \frac{1}{33} \sigma_1^2 \sigma_2^2 +\frac{6}{11} \sigma_1\sigma_2\sigma_3 +\frac{16}{231} \sigma_2^3 +\sigma_3^2\,,\\
\underline{s=8}:&&
\sigma_2^\sevenh A_{2,1}=\frac{3}{143}\sigma_1^3 \sigma_2^2 +\frac{28}{429}\sigma_1 \sigma_2^3 -\frac{6}{13}\sigma_1^2 \sigma_2\sigma_3 +\sigma_1\sigma_3^2-\frac{4}{143} \sigma_2^2\sigma_3\,,\\*&&
\sigma_2^\sevenh A_{1,4}=\frac{1}{13}\sigma_1^5\sigma_2 {-}\sigma_1^4\sigma_3 {+}\frac{2}{143}\sigma_1^3\sigma_2^2{+}\frac{18}{13}\sigma_1^2\sigma_2\sigma_3{-}\frac{43}{429}\sigma_1\sigma_2^3 {-} \frac{27}{143}\sigma_2^2\sigma_3\,,\\
\underline{s=10}:&&
\sigma_2^\nineh A_{3,0}=\frac{1}{221}\sigma_1^3 \sigma_2^3-\frac{3}{17} \sigma_1^2\sigma_2^2\sigma_3 +\frac{72}{2431}\sigma_1\sigma_2^4 +\frac{15}{17}\sigma_1\sigma _2\sigma_3^2 \nn \\*  && \qquad\quad\phantom{=}
 -\frac{24}{221}\sigma_2^3 \sigma_3-\sigma_3^3\,,\\ &&
\sigma_2^\nineh  A_{2,3}=\frac{1}{85} \sigma_1^5 \sigma_2^2 -\frac{6}{17} \sigma_1^4 \sigma_2 \sigma_3+\frac{11}{221}\sigma_1^3 \sigma_2^3 +\sigma_1^3 \sigma_3^2 +\frac{2}{17}\sigma_1^2 \sigma_2^2 \sigma_3 \nn\\* &&\qquad\quad \phantom{=}-\frac{4}{143}\sigma_1 \sigma_2^4 
 -\frac{9}{17}\sigma_1\sigma_2\sigma_3^2+\frac{4}{221}\sigma_2^3 \sigma_3\, .
\label{eq:aijsome}
\end{eqnarray}}


\section{From lattice sums to generalised Eisenstein series}
\label{app:LapEig}
In order to determine the properties of the modular invariant functions which can be represented via the lattice sum integrals just discussed, it is very useful to apply the laplacian $\Delta_\stau$  to  \eqref{eq:defEwijApp} which gives
\begin{align}  
\Delta_\tau  \cE^w_{i,j}(\stau) =   \sum_{\underset {p_1+p_2+p_3=0} {p_1,p_2, p_3\ne 0}} \int_0^\infty d^3 t\, [V(t)]^{w-3}  \, A_{i,j}(\ttau(t))\, \Delta_t \exp\Big( - \frac{\pi}{\tau_2}\sum_{i=1}^3 t_i |p_i|^2  \Big)\,,
\label{eq:defEwij2}
\end{align}
where $\Delta_\ttau = \ttau_2^2 (\partial_{\ttau_1}^2 +\partial_{\ttau_2}^2)$, and we have used  \eqref{eq:DeltaTauOm}.  We now integrate by parts and note
the property 
\begin{equation}
\Delta_t [V(t)^\alpha F(t)] = V(t)^\alpha \Delta_t [ F(t)] \, , \label{eq:deltatV}
\end{equation}
for all $\alpha$.
We will also define
\begin{equation}\label{eq:Bwij}
B^{w}_{i,j}(t) \equiv[V(t)]^{w-3} A_{i,j}(\ttau(t))\,,
\end{equation}
and note that \eqref{DeltaA} and \eqref{eq:deltatV} imply
\begin{equation}
\Delta_t\Big[ B^{w}_{i,j}(t)\Big] =  [V(t)]^{w-3} \Big[\Delta_\ttau A_{i,j}(\ttau)\Big]_{\ttau=\ttau(t)} = s(s-1) B^{w}_{i,j}(t)\, , 
\end{equation}
with $s=3i+j+1$. It follows that 
\begin{align}
\Delta_\stau \cE^w_{i,j}(\stau) &\nn=  \sum_{\underset {p_1+p_2+p_3=0} {p_1,p_2, p_3\ne 0}} \int_0^\infty d^3 t\, B^{w}_{i,j}(t)\, \Delta_t\Big[ \exp\Big( - \frac{\pi}{\tau_2}\sum_{i=1}^3 t_i |p_i|^2  \Big) \Big]\\
&\nn=  \sum_{\underset {p_1+p_2+p_3=0} {p_1,p_2, p_3\ne 0}} \int_0^\infty d^3 t \, \Delta_t\big[B^{w}_{i,j}(t)\big]\,  \exp\Big( - \frac{\pi}{\tau_2}\sum_{i=1}^3 t_i |p_i|^2  \Big) +{\rm{\bt}}\\
& = s(s-1)  \cE^w_{i,j}(\stau) +{\rm{\bt}}\, ,\label{eq:LapIn}
\end{align}
where `${\rm{\bt}}$' represents  boundary terms. 

The  boundary terms  in \eqref{eq:LapIn} are easily collected by integrating $\Delta_t$ by parts using the definition \eqref{eq:deltat}.
We start by noting that since all the momenta $p_1,p_2,p_3=-p_1-p_2$ are non-vanishing we only need to worry about boundary contributions coming from $t_i\to0$ since the integral is exponentially suppressed along any direction as $t_i\to \infty$.
Using the fact that the integrand is invariant under permutations of $(t_1,t_2,t_3)$ we arrive at the expression
\begin{align}
{\rm{\bt}} = &\nn\sum_{\underset {p_1+p_2+p_3=0} {p_1,p_2, p_3\ne 0}} 3 \int_0^\infty d^2t \,\Big[t_1t_2(\partial_3 -\partial_1-\partial_2) B^{w}_{i,j}(t_1,t_2,t_3)\Big]_{t_3=0} \exp\Big(\! -\frac{ \pi t_1 |p_1|^2}{\tau_2}-\frac{ \pi t_2 |p_2|^2}{\tau_2}\Big)\\
&+\sum_{p\neq0 } \frac{6\pi |p|^2}{\tau_2} \int_0^\infty d^2t \,   \Big[t_1t_2B^{w}_{i,j}(t_1,t_2,t_3)\Big]_{t_3=0} \exp\Big(-\frac{\pi |p|^2 (t_1+t_2)}{\tau_2}\Big)\,,\label{eq:BT}
\end{align}
where 
we have defined $p=m+n \tau$ with $(m,n) \in \mathbb{Z}^2 \slash \{ (0,0)\}$.
It is easy to check with the definition of $A_{i,j}$ given above that the first term produces {terms} bilinear in Eisenstein series, while the second term will produce {terms}  linear in Eisenstein series.

Although we do not have a closed formula for the boundary term associated with a specific $B^{w}_{i,j}$, we can make some general observations.
It follows from  \eqref{Aijexp} that when $A_{i,j}(\ttau)$ is rewritten in terms of  $t_i$  and expressed in terms of  the symmetric variables $\sigma_1,\sigma_2,\sigma_3$ defined in \eqref{eq:symVar}, it  takes the general form
\begin{equation}\label{eq:eq:Aijexp}
A_{i,j}(\ttau(t)) = \sum_{\alpha=0}^{s-1}\sum_{\beta = \frac{s \,{\rm mod }\,2}{2} }^{\frac{s-1}{2}} \sum_{\gamma =0}^{\lfloor \frac{s-1}{3}\rfloor} c(\alpha,\beta,\gamma) \sigma_1^\alpha  \sigma_2^{-\beta}  \sigma_3^\gamma\,,
\end{equation}  
 where the coefficients $c(\alpha,\beta,\gamma)\in \mathbb{Q}$ implicitly depend on $i,j$ (and  $s=3i+j+1$).

We note that when $s$ is even $\beta \in \mathbb{N}$ while for $s$ odd $\beta\in \mathbb{N}+\frac{1}{2}$. Furthermore the $A_{i,j}$ are homogeneous functions of $t_i$ since $\ttau_1$ and $\ttau_2$ (defined in \eqref {eq:rhodef}) are, and  hence it follows that 
\begin{equation}
\alpha -2\beta +3 \gamma=0\,.
\end{equation}
From  \eqref {eq:rhodef} we also deduce that the transformation $\ttau\to \bar\ttau$ is equivalent to $t_i\to -t_i$, or equivalently 
\ie
\sigma_1\to -\sigma_1\, , \qquad \sigma_2\to \sigma_2\, , \qquad \sigma_3 \to -\sigma_3 \, .
\fe 
Hence, using \eqref{Aijexp}, we see that 
\begin{align}
&\nn A_{i,j}(\bar\ttau)  = (-1)^{i+j} A_{i,j}(\ttau) = (-1)^{s-1} A_{i,j}(\ttau)\\
 \Longrightarrow \qquad &A_{i,j}(\ttau(-t)) = A_{i,j}(\bar{\ttau}(t)) = (-1)^{s-1} A_{i,j}(\ttau(t))\,,
\end{align}
which from the expansion \eqref{eq:eq:Aijexp} implies 
\begin{equation}
c(\alpha,\beta,\gamma) = 0 \, ,  \qquad {\rm for}\qquad \alpha+\gamma \not\equiv (s-1)\, ~ {\rm mod}\, ~ 2\,. \label{eq:condmod2}
\end{equation}

We can then focus on the integrand \eqref{eq:Bwij} and expand it as
\begin{equation}
B^{w}_{i,j}(t_1,t_2,t_3) = \sum_{\alpha=0}^{s-1}\sum_{\beta = \frac{s \,{\rm mod }\,2}{2} }^{\frac{s-1}{2}} \sum_{\gamma =0}^{\lfloor \frac{s-1}{3}\rfloor} c(\alpha,\beta,\gamma)\, \sigma_1^\alpha \,\sigma_2^{\frac{w-3}{2}-\beta}\sigma_3^\gamma\,,
\end{equation}
from which it is rather easy to compute the boundary contribution \eqref{eq:BT} where only the terms with $\gamma=0$ and $\gamma=1$ contribute.
The boundary terms are given by
\begin{align}
{\rm{\bt}} = &\nn \, -\!\!\!\! \sum_{\alpha=(s-1)\, {\rm mod}\,2}^{s-1} \,\,\sum_{\underset {p_1+p_2+p_3=0} {p_1,p_2, p_3\ne 0}}  3 \alpha c(\alpha,\tfrac{\alpha}{2},0) \int_0^\infty d^2t \,(t_1 t_2)^{\frac{w-1-\alpha}{2}}(t_1+t_2)^{\alpha-1}  e^{ -\frac{ \pi t_1 |p_1|^2}{\tau_2}-\frac{ \pi t_2 |p_2|^2}{\tau_2}}\\
&+\nn\sum_{\alpha=s \,{\rm mod}\,2}^{s-2}\,\, \sum_{\underset {p_1+p_2+p_3=0} {p_1,p_2, p_3\ne 0}} 3  c(\alpha,\tfrac{\alpha+3}{2},1) \int_0^\infty d^2t \, (t_1 t_2)^{\frac{w-2-\alpha}{2}}(t_1+t_2)^{\alpha}  e^{-\frac{ \pi t_1 |p_1|^2}{\tau_2}-\frac{ \pi t_2 |p_2|^2}{\tau_2}}\\
&+\sum_{\alpha=(s-1)\, {\rm mod}\,2}^{s-1} \,\,\sum_{p\neq 0 } \frac{6\pi |p|^2}{\tau_2}    c(\alpha,\tfrac{\alpha}{2},0)\int_0^\infty d^2t \,(t_1 t_2)^{\frac{w-1-\alpha}{2}}(t_1+t_2)^{\alpha} e^{-\frac{\pi |p|^2 (t_1+t_2)}{\tau_2}}\,.\label{eq:BT1}
\end{align}
Note that the sum truncates at  $\alpha=s-2$ in the second line  since \eqref{eq:condmod2} implies that  $c(s-1,\beta,1)=0$  for $\alpha =s-1$.

Since $\alpha\in \mathbb{N}$ we can use the binomial expansion on $(t_1+t_2)^\alpha$ and integrate term by term, arriving at the sum of three boundary contributions, 
\begin{align}
&\nn{\rm{\bt}} = \\
&\nn -3\! \sum_{\alpha=(s-1)\, {\rm mod}\,2}^{s-1} \sum_{\delta=\frac{1-\alpha}{2}}^{\frac{\alpha-1}{2}} \alpha {\alpha{-}1 \choose \frac{\alpha-1}{2}{+}\delta } c(\alpha,\tfrac{\alpha}{2},0)\Gamma\big(\tfrac{w}{2}{+}\delta\big)\Gamma\big(\tfrac{w}{2}{-}\delta\big) \Big[E(\tfrac{w}{2}{+}\delta;\tau) E(\tfrac{w}{2}{-}\delta;\tau){-}E(w;\tau)\Big]\\
&\nn + 3\! \sum_{\alpha=s \,{\rm mod}\,2}^{s-2}\, \sum_{\delta=\frac{-\alpha}{2}}^{\frac{\alpha}{2}} {\alpha \choose \frac{\alpha}{2}{+}\delta} c(\alpha,\tfrac{\alpha+3}{2},1)\Gamma\big(\tfrac{w}{2}{+}\delta\big)\Gamma\big(\tfrac{w}{2}{-}\delta\big) \Big[E(\tfrac{w}{2}{+}\delta;\tau) E(\tfrac{w}{2}{-}\delta;\tau){-}E(w;\tau)\Big]\\
& +6 \sum_{\alpha=(s-1)\, {\rm mod}\,2}^{s-1}\, \sum_{\delta=\frac{1-\alpha}{2}}^{\frac{\alpha+1}{2}}  {\alpha-1 \choose \frac{\alpha-1}{2}+\delta } c(\alpha,\tfrac{\alpha}{2},0)\Gamma\big(\tfrac{w}{2}+\delta\big)\Gamma\big(\tfrac{w}{2}+1-\delta\big) E(w;\tau)\,,\label{eq:sources}
\end{align}
 where we have used the identity
\begin{align}
\sum_{\underset {p_1+p_2+p_3=0} {p_1,p_2, p_3\ne 0}} \frac{(\tau_2/\pi)^{s_1+s_2}}{ |p_1|^{2s_1} |p_2|^{2s_2}} &= \sum_{p_1,p_2\neq0 }  \frac{(\tau_2/\pi)^{s_1+s_2}}{ |p_1|^{2s_1} |p_2|^{2s_2}} - \sum_{p\in \Lambda'}  \frac{(\tau_2/\pi)^{s_1+s_2}}{ |p|^{2(s_1+s_2)}} \cr
&= E(s_1;\tau)E(s_2;\tau)- E(s_1+s_2;\tau)\,.
\end{align}

Note that if the eigenvalue $s$ in  \eqref{eq:sources} is even then the sum in the first line is over odd values of $\alpha$ since otherwise $c(\alpha,\beta,0)$ vanishes from \eqref{eq:condmod2}, so $\delta$ must be an integer. Similarly, if $s$ is even the $\alpha$ sum in the second line is over even values since  otherwise  $c(\alpha,\beta,1)$ vanishes  again from \eqref{eq:condmod2}, and once more $\delta$ is an integer. 
Conversely for $s$ odd in the first line the sum over $\alpha$ is over even integers, so  $\delta$ is half-integer.  For $s$ odd in the second line the  $\alpha$ sum is over odd values and  the  $\delta$ sum is again over half-integers. We conclude that for even eigenvalue $s$ the bilinears terms in both sums are  of the form $E(\frac{w}{2}+\delta;\tau) E(\frac{w}{2}-\delta;\tau)$ with integer $\delta$.   The third line gives terms proportional to non-holomorphic Eisenstein series with integer indices. Thus we have reproduce the claim \eqref{eq:LapcEijw} stated in the main text.

\subsubsection*{Comparison with modular graph functions}

In the case of modular graph functions even eigenvalue $s$ is accompanied by an even trascendental weight $w\geq 4$ such that $2\leq s\leq w-2$, hence in this case the sources are always of the form $E(s_1;\tau) E(s_2;\tau)$ with $s_1,s_2\in \mathbb{N}^{\geq 2}$.
Conversely, for the generalised Eisenstein of present interest even eigenvalue $s$ is accompanied by an odd trascendental weight $w$ such that $s\geq w+1$. In this case the bilinear sources are always of the form $E(s_1;\tau) E(s_2;\tau)$ with $s_1,s_2 \in \mathbb{Z}+\frac{1}{2}$ such that $s_1+s_2=w$.

Mutatis mutandis, modular graph functions with odd eigenvalue $s$ must have odd trascendental weight $w\geq 5$ such that $3\leq s\leq w-2$, hence again the sources are always of the form $E(s_1;\tau) E(s_2;\tau)$ with $s_1,s_2\in \mathbb{N}^{\geq 2}$. While again the generalised Eisenstein  series' of present interest and odd eigenvalue $s$ are accompanied by an even weight $w$ and again  the bilinear sources are of the form $E(s_1;\tau) E(s_2;\tau)$ with $s_1,s_2 \in \mathbb{Z}+\frac{1}{2}$ such that $s_1+s_2=w$.

\bibliographystyle{JHEP}
\bibliography{Second}

\end{document}